%% file: main.tex
\documentclass[acmsmall]{acmart}

\AtBeginDocument{%
  \providecommand\BibTeX{{%
    \normalfont B\kern-0.5em{\scshape i\kern-0.25em b}\kern-0.8em\TeX}}}

\setcopyright{rightsretained}
\acmJournal{PACMMOD}
\acmYear{2023} \acmVolume{1} \acmNumber{1} \acmArticle{30} \acmMonth{5} \acmPrice{}
\acmDOI{10.1145/3588710}
\acmISBN{2836-6573/2023/5 - 30}

\usepackage[utf8]{inputenc} 
\usepackage[T1]{fontenc}    
\usepackage{hyperref}       
\usepackage{url}            
\usepackage{booktabs}       
\usepackage{amsfonts}       
\usepackage{nicefrac}       
\usepackage{microtype}      
\usepackage{xcolor}         
\usepackage{amsmath}
\usepackage{graphicx}
\usepackage{balance}
\usepackage{subcaption}
\usepackage{caption}
\usepackage{fancybox}
\usepackage{todonotes}
\usepackage{threeparttable}
\setuptodonotes{inline}
\usepackage{svg}
\usepackage{multicol, multirow}
\usepackage{makecell}
\usepackage{comment}
\usepackage[normalem]{ulem}
\usepackage{tablefootnote}
\usepackage{enumitem}
\usepackage[boxed,linesnumbered]{algorithm2e}
\usepackage[font={small}]{caption}

\SetAlCapSkip{1em}

\newif\ifnotes
\notestrue

\newcommand{\ignore}[1]{}
\renewcommand{\paragraph}[1]{\vspace{0.5\baselineskip}\noindent\textbf{#1.}\hspace{0.1cm}}
\newcommand{\code}[1]{\texttt{#1}}
\newcommand{\gittables}{{\sc GitTables}\xspace}

\begin{document}

\title{GitTables: A Large-Scale Corpus of Relational Tables}

\author{\href{https://orcid.org/0000-0002-0949-7290}{Madelon Hulsebos}}
\authornote{\vspace{-.1cm}Corresponding author (\href{mailto:m.hulsebos@uva.nl}{m.hulsebos@uva.nl})\vspace{-.1cm}}
\affiliation{
  \institution{University of Amsterdam}
  \city{Amsterdam}
  \postcode{1012 WX}
  \country{Netherlands}
}
\email{m.hulsebos@uva.nl}

\author{\href{https://orcid.org/0009-0003-2080-0443}{{\c{C}}a{\u{g}}atay Demiralp}}
\affiliation{
  \institution{Sigma Computing}
  \city{San Francisco}
  \postcode{CA 94105}
  \country{United States}
}
\email{cagatay@sigmacomputing.com}

\author{\href{https://orcid.org/0000-0003-0183-6910}{Paul Groth}}
\affiliation{%
  \institution{University of Amsterdam}
  \city{Amsterdam}
  \postcode{1012 WX}
  \country{Netherlands}
}
\email{p.t.groth@uva.nl}

\input{00_abstract}

\begin{teaserfigure}
    \centering
    \includegraphics[width=\textwidth]{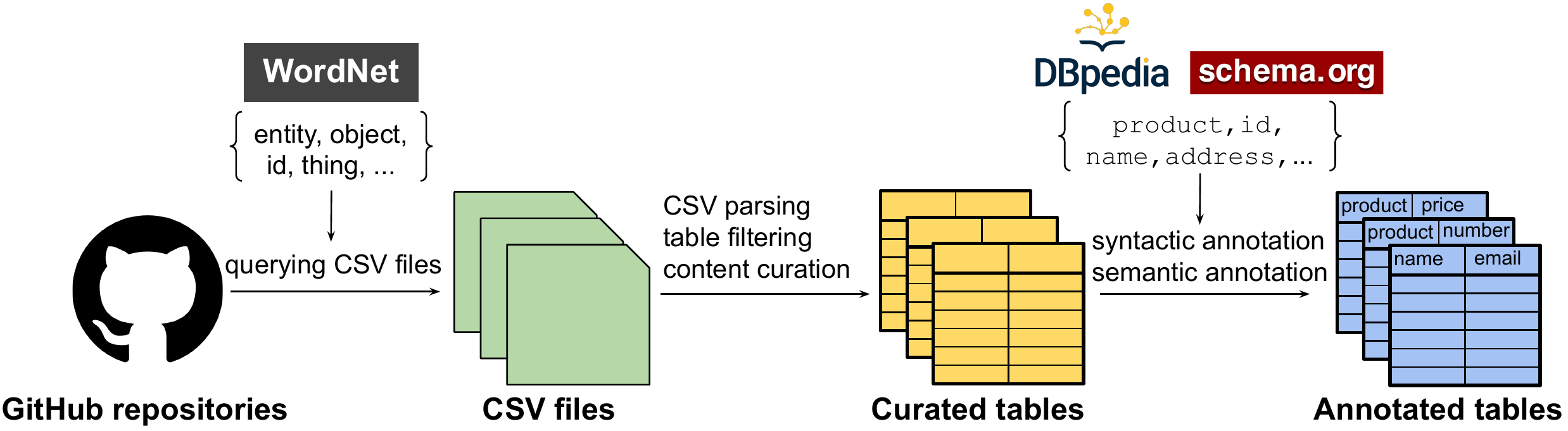}
    \caption{The pipeline for creating \gittables consists of 1) querying CSV files from GitHub repositories based on WordNet, 2) parsing CSV files and curating tables, and 3) annotating tables with column semantics from DBpedia and Schema.org.}
    \label{fig:gittables-pipeline}
\end{teaserfigure}



\maketitle

\input{01_introduction}

\input{02_related}

\input{03_gittables}

\input{04_analysis}

\input{05_applications}

\input{07_conclusion}

\balance

\bibliographystyle{ACM-Reference-Format}
\bibliography{refs}

\received{April 2022}
\received[accepted]{July 2022}

\end{document}

%% file: 00_abstract.tex
\begin{abstract}

The success of deep learning has sparked interest in improving relational table tasks, like data preparation and search, with table representation models trained on large table corpora. Existing table corpora primarily contain tables extracted from HTML pages, limiting the capability to represent offline database tables. To train and evaluate high-capacity models for applications beyond the Web, we need resources with tables that resemble relational database tables.
Here we introduce \gittables, a corpus of 1M relational tables extracted from GitHub. Our continuing curation aims at growing the corpus to at least 10M tables. 
Analyses of \gittables show that its structure, content, and topical coverage differ significantly from existing table corpora. We annotate table columns in \gittables with semantic types, hierarchical relations and descriptions from Schema.org and DBpedia. The evaluation of our annotation pipeline on the T2Dv2 benchmark illustrates that our approach provides results on par with human annotations.
We present three applications of \gittables, demonstrating its value for learned semantic type detection models, schema completion methods, and benchmarks for table-to-KG matching, data search, and preparation. We make the corpus and code available at \url{https://gittables.github.io}.

\end{abstract}

%% file: 01_introduction.tex
\section{Introduction}\label{sec:intro}

Deep learning (DL) models, in the past decade, have dramatically improved many long-standing computer vision and natural language processing tasks~\cite{lecun2015deep}. The practical success of DL has also spurred interest in its applications to tasks in other domains, including data management, ranging from data cleaning to annotation~\cite{wang2016database}. To train DL models for relational data, earlier work primarily relied on corpora consisting of tables scraped from HTML pages~\cite{cafarella2008uncovering} such as WDC WebTables~\cite{webtables2012}, the largest table corpus to date. These models have played an important role in facilitating data-driven data management research, with Web applications in particular~\cite{webtablestenyears, zhang2020web}.

However, tables extracted from HTML pages on the Web (Web tables) provide a skewed representation of tables in the wild especially those residing in (enterprise) databases~\cite{li2017discoveringenterprise, langenecker2021towards}. The common attribute ``ID'', for example, does not appear among the twenty most frequent column names of the WDC WebTables corpus, and the dimensions of Web tables are significantly smaller than typical database tables~\cite{webtables2012}. It is therefore unsurprising that models based on Web tables have limited applications beyond the Web~\cite{webtablestenyears}. To broaden the impact of data-driven data management research, we need new table collections complementing existing corpora with tables resembling typical database tables.

To address this demand, we introduce \gittables: a corpus with 1M relational tables extracted from CSV files in GitHub repositories. We will keep the corpus growing to have at least 10M tables to facilitate the extension of deep transfer learning to the relational domain, analogous to how large corpora of natural language have stimulated pretrained language models like BERT~\cite{devlin2019bert} and GPT~\cite{brown2020gpt3}. We expect \gittables to stimulate similar progress in data management tasks like data search and preparation.

Besides having representative tables, many data management applications benefit from understanding the semantics of table columns. An intelligent data exploration system, for example, could recommend a map chart to visualize two data columns representing countries and sales revenues. Upon encountering missing values, the system could also automatically fill the gaps using a knowledge base. The system could, for instance, impute ``France'' in an empty cell in a column with country values, if the neighboring cell in a column with capital cities is ``Paris''.

\begin{figure}
    \centering
    \vspace{-0.5cm}
    \includegraphics[width=\columnwidth]{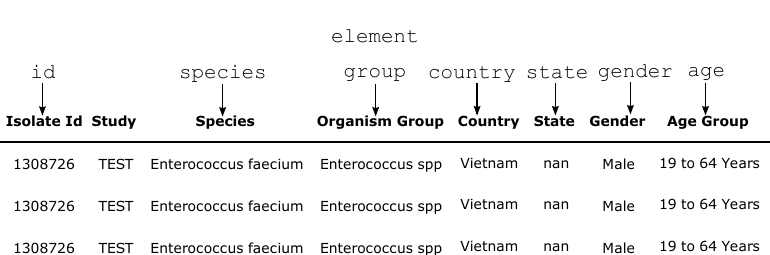}
    \vspace{-0.5cm}
    \caption{Snippet of an annotated table in \gittables. Annotations are provided with confidence scores. For example, the confidence of the type \code{species} for the column ``Species'' is 1, while \code{element group} for ``Organism Group'' is 0.7. These scores enable filtering based on the intended use case.}
    \label{fig:annotation-example}
\end{figure}

To facilitate such functionalities, we enhance table columns with semantic types from DBpedia and Schema.org (e.g. \code{address}) using a syntactic as well as semantic annotation method (an example in Figure~\ref{fig:annotation-example}). 
Each semantic type is also associated with the expected atomic data type (e.g. \code{string} and \code{number}), a description, and hierarchical type relations. These semantically rich annotations together with DL present a unique opportunity to learn table semantics as demonstrated with TURL~\cite{turl} and TaBERT~\cite{yin2020tabert}.

Our analysis of \gittables confirms the different nature of the tables: the tables have significantly larger dimensions (rows and columns). A machine learning model for detecting data shift between \gittables and Web tables accurately distinguishes table columns from each source corpus. This reflects the structurally different content of these corpora. The semantic type distribution also deviates significantly from the semantic distribution of Web tables, illustrating its distinctive topical coverage. We make \gittables available through \url{https://gittables.github.io}. We contribute:
\begin{enumerate}[leftmargin=0.5cm]
    \vspace{-0.15cm}
    \item \gittables, a new large-scale corpus of 1M tables. To the best of our knowledge, \gittables is the first large-scale relational table corpus with topical coverage and content structurally different than tables extracted from HTML pages.
    \item A scalable column-annotation method using distant-supervision. We annotate the columns in \gittables with semantics consisting of semantic types, atomic data types, hierarchical relations, and descriptions, making \gittables the largest annotated table corpus to date.
    \item Three applications demonstrating the value of \gittables: 1) we train a semantic type detection model and obtain high prediction accuracy, 2) we use \gittables as a resource for schema completion methods in typical database contexts, and 3) we present a starting point for benchmark datasets for typical data management tasks.
\end{enumerate}

%% file: 02_related.tex
\begin{table*}
  \caption{Existing large-scale relational table corpora in comparison with \gittables which has considerably higher dimensions, resembling actual tables in databases. We report statistics related to relational tables, leaving out, for example, entity tables.}
  \vspace{-0.2cm}
  \label{tab:table-corpora}
  \centering
  \small
  \begin{threeparttable}
      \begin{tabular}{llccc}
        \toprule
        \setlength\tabcolsep{1.5pt}
        \textbf{Name} & \textbf{Table source} & \textbf{\# tables} & \textbf{Avg \# rows} & \textbf{Avg \# cols} \\
        \midrule
        WDC WebTables \cite{Lehmberg:2016:WebTableCorpus}  & HTML pages & 90M & 11 & 4 \\
        Dresden Web Table Corpus \cite{dresdenwebtables} & HTML pages & 59M & 17 & 6 \\
        WikiTables \cite{bhagavatula2013wikitables} & Wikipedia tables & 2M & 15 & 6\\
        Open Data Portal Watch \ \cite{mitlohner2016characteristics} & CSVs from Open Data portals & 107K & 365 & 14  \\
        VizNet \cite{hu2019viznet} & WebTables, Plotly, i.a. & 31M & 17 & 3  \\
        \textbf{GitTables} & \textbf{CSVs from GitHub} & \textbf{1M} & \textbf{142} & \textbf{12}  \\
        \bottomrule
      \end{tabular}
    \end{threeparttable}
    \normalsize
    \vspace{-.5cm}
\end{table*}

\section{Related work}\label{sec:related-work} 
Web initiatives such as Common Crawl, Wikipedia, and Open Data have been cost-effective resources for curating unstructured and structured data at scale~\cite{muhleisen2012web, bhagavatula2013wikitables, neumaier2016automated}. Below we discuss large-scale table corpora sourced from these initiatives and review prior work annotating column semantics of tables sampled from these corpora.

\subsection{Large-scale table corpora} \label{sec:table-corpora}
\vspace{-0.1cm}
\paragraph{WDC WebTables and Dresden Web Table Corpus}
These Web table corpora~\cite{Lehmberg:2016:WebTableCorpus,dresdenwebtables} extract tables from HTML pages in the Common Crawl corpus as inspired by~\cite{carafella2008webtables}. They provide an abundance of relational tables ranging from 59M to 90M and have been instrumental in advancing applications like table augmentation and integration~\cite{webtablestenyears, zhang2020web}. However, Web tables generalize poorly due to their small dimensions and different content~\cite{langenecker2021towards}.

\paragraph{WikiTables}
To provide high-quality tables with semantics that are easier to detect than those of arbitrary Web tables, WikiTables extracts approximately 2M tables from Wikipedia~\cite{bhagavatula2013wikitables}. This corpus is primarily suitable for tasks such as question answering that rely on the quality of the table contents. Unsurprisingly, the tables in WikiTables are just as small as those in WDC WebTables.

\paragraph{Open Data Portal Watch}
With 227K CSV files extracted from 260 Open Data portals~\cite{neumaier2016automated}, this is the first substantial corpus not sourced from HTML pages. 108K of these CSV files were parsed to tables and analysed based on their format, structure, and data type~\cite{mitlohner2016characteristics}. This analysis illustrates the different table dimensions and atomic data type distributions of such tables compared to Web table corpora, motivating the construction of a corpus like \gittables.

\paragraph{VizNet}
VizNet was constructed to train and evaluate visualization methods with real-world tables~\cite{hu2019viznet}. It combines 31M tables from WebTables ~\cite{carafella2008webtables}, ManyEyes~\cite{viegas2007manyeyes}, Plotly~\cite{plotlyfeed}, and Open Data portals~\cite{neumaier2016automated}. Analyses of VizNet suggest that tables not from the Web exhibit different internal structures, providing further evidence for differences between Web tables and tables from other sources.

\begin{table*}
  \vspace{.8cm}
  \caption{Characteristics of annotated relational table datasets. Existing annotated corpora are limited in the number of (relational) tables and types, while \gittables provides a large-scale corpus annotated with over 2K semantic types.}
  \label{tab:annotated-table-corpora}
  \centering
  \small
  \begin{threeparttable}
      \begin{tabular}{llccccl}
        \toprule
        \textbf{Dataset} & \textbf{Table source} & \textbf{\# tables} & \textbf{Avg \# rows} & \textbf{Avg \# cols} & \textbf{\# types} & \textbf{Ontology} \\
        \midrule
        T2Dv2 \cite{t2dv2} & WebTables & 779 & 17 & 4 & 275 & DBpedia \\
        SemTab\tnote{1} \ \cite{jimenez2020semtab} & WikiData, Wikipedia & 132K & 224 & 4 & - & DBpedia \\
        TURL \cite{turl} & WikiTables & 407K & 18 & 3 & 255 & Freebase \\
        \multirow{2}{*}{\textbf{GitTables}} & \multirow{2}{*}{\textbf{CSV files on GitHub}} & \multirow{2}{*}{\textbf{962K}} & \multirow{2}{*}{\textbf{142}} & \multirow{2}{*}{\textbf{12}} & \multirow{2}{*}{\textbf{2.4K}} & \textbf{DBpedia} \\
        & & & & & & \textbf{Schema.org} \\
        \bottomrule
      \end{tabular}
        \begin{tablenotes}
        \item[1]Statistics aggregated over all datasets included in the challenge.
        \end{tablenotes}
    \end{threeparttable}
    \normalsize
\end{table*}

\subsection{Table datasets with column annotations} \label{sec:annotated-table-corpora}
\paragraph{T2Dv2} This is a subset of WDC WebTables and was curated for benchmarking methods for knowledge base (KB) augmentation~\cite{ritze2017matching}. The rows, columns and tables were manually annotated with correspondences to DBpedia instances, properties and classes. This was found to be a trivial target for KB matching due to the many ``obviously'' linkable entities. Recent work also points out that Web tables might not be typical of tables used for KB augmentation \cite{cutrona2020tough}.

\paragraph{SemTab} The SemTab challenge~\cite{jimenez2020semtab} provides datasets for benchmarking KG matching methods. Most of the tables were extracted from Wikidata and ``refined'' by adding for example noise. SemTab 2020 incorporates 180 large-sized Wikipedia tables which enriched with noise to mimic real tables~\cite{cutrona2020tough}. Columns were annotated by linking column values to DBpedia types and aggregating these types to a column-level annotation. \gittables is a useful resource in future versions of the SemTab challenge for benchmarking table interpretation methods on database-like tables.

\paragraph{TURL} Inspired by pretrained language models (e.g.,~\cite{devlin2019bert,raffel2020t5,brown2020gpt3}), TURL~\cite{turl} provides a framework for learning embedding representations of Web tables through a pretrained model. Pretrained models require fine-tuning with labeled data to be applied to specific tasks in domains. To do this, a set of tables from WikiTables is annotated with 255 semantic types from Freebase. Application of learned table representations for table understanding motivates the construction of a larger-scale and rich corpus to support this nascent line of research.

%% file: 03_gittables.tex
\section{GitTables}
In this section, we summarize the design principles behind \gittables and describe the construction pipeline in detail.

\vspace{0.2cm}

\subsection{Design principles of GitTables}

Based on the gaps reported in the literature and our experience developing learned models for table interpretation tasks, we identified four criteria for \gittables:

\begin{itemize}[leftmargin=0.55cm, rightmargin=0em, itemsep=0.2em]
 \item[\textbf{C1}] To facilitate learned table representations that generalize across different database contexts, the scale of the corpus should largely exceed the scale of the Open Data portals.
 \item[\textbf{C2}] To advance research beyond the Web, we need ``database''-like tables which are considerably larger and more heterogeneous than typical Web tables.
 \item[\textbf{C3}] The corpus should have topical coverage and content that generalizes to enterprises, governments, and beyond.
 \item[\textbf{C4}] Tables should be enriched with semantic annotations to facilitate the development of learned models for data management tasks like data validation and preparation.
\end{itemize}

We considered different public interfaces to retrieve structured data files for extracting relational tables, like Zenodo and. Kaggle. On GitHub, we then performed a simple search for CSV files, yielding 92,191,141 files\footnote{\url{https://github.com/search?q=extension\%3Acsv\&type=Code} (3 February 2022)}. This suggests that GitHub can be an effective resource for collecting a relational table corpus at scale (criterion \textbf{C1}). We focused on the CSV format given its wide use for storing raw structured data~\cite{mitlohner2016characteristics}, as reflected by the massive number of CSV files we found in our search.

\begin{figure}[!]
    \centering
    \shadowbox{\includegraphics[width=0.96\columnwidth]{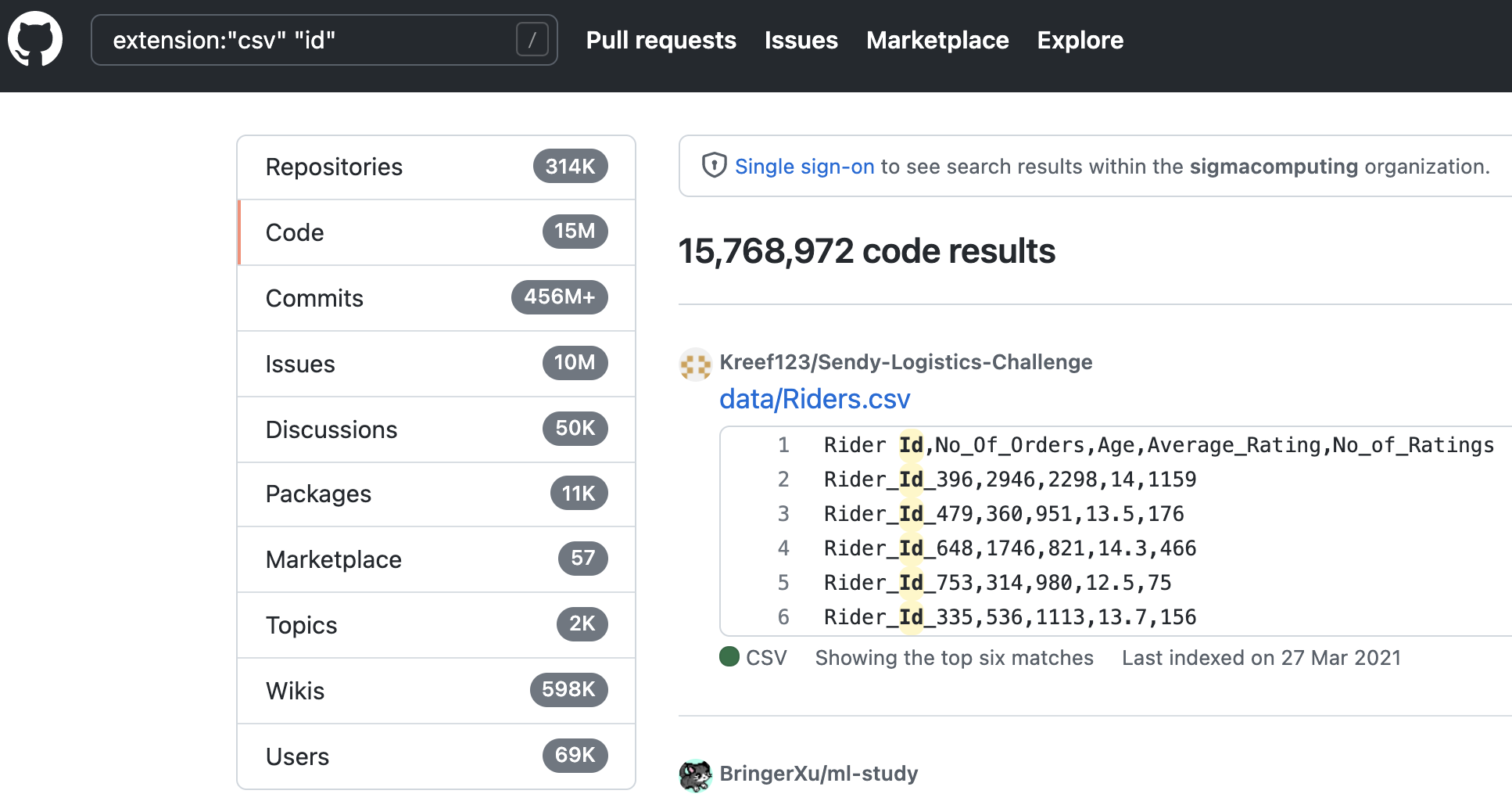}}
    \caption{The results when querying for CSV files with the term ``id'' as shown on GitHub its website. It illustrates the scale of CSV files (15M+ files for this particular query) and the database-like tables that are stored on GitHub.}
    \label{fig:github-query-example}
\end{figure}

GitHub is commonly used by programmers, data scientists, and researchers, among others~\cite{perkel2016democratic, koesten2020dataset}. Given the proximity of such users to actual databases, GitHub is a rich source for heterogeneous tables. Figure ~\ref{fig:github-query-example} shows the first CSV file retrieved by GitHub when it is queried for the term ``id''. This table with data about ``riders`` shows column names and table contents that are typical for database tables. This figure also illustrates the abundance of similar CSV files present on GitHub, as we can find close to 16M CSV files for this single query only.
Prior analyses of CSV files from GitHub also found that these files have diverse formatting and the tables extracted from them have relatively large dimensions~\cite{van2019wrangling, koesten2020dataset}, which is common across database contexts~\cite{van2019wrangling, langenecker2021towards}. Altogether, we consider CSV files from GitHub a suitable resource for database-like tables (\textbf{C2}).

Inspired by the construction of ImageNet~\cite{imagenet}, we select 67K unique English nouns from WordNet~\cite{wordnet} yielding a set of diverse keywords (called ``topics'') to specify our search queries. Although it introduces some bias towards English tables, WordNet nouns have the desired topical coverage to ensure content diversity in \gittables (\textbf{C3}). To avoid the ``WordNet effect''~\cite{birhane2021large}, we remove topics like ``killing'' which might yield tables with offensive content and are out of scope.

To satisfy criterion \textbf{C4}, we provide semantic annotations for table columns. We developed two annotation methods, one syntactic method informed by Sherlock~\cite{hulsebos2019sherlock} and the other leveraging a pretrained semantic model. We considered multiple ontologies to accommodate different use cases, and selected DBpedia~\cite{dbpedia} and Schema.org~\cite{guha2016schema} as they are well-curated and provide complementary and diverse semantic types (as discussed further in Section~\ref{sec:annotating-columns}).

Altogether, the high-level pipeline for constructing \gittables comes down to 1) extracting CSV files from GitHub, 2) parsing and curating tables from CSV files, and 3) annotating tables with column semantics. Figure \ref{fig:gittables-pipeline} visualizes this pipeline.

\subsection{Extracting CSV files from GitHub} \label{sec:csv-extraction}

GitHub restricts querying in multiple ways to avoid overloading its Search API. First, it is not possible to retrieve files larger than 438 kB which bounds the CSV files we extract in terms of their size. Although some organizations may use larger files, most CSV files in Open Data portals are found to be smaller than 100 kB~\cite{mitlohner2016characteristics}. We also observed repositories with larger tables that were split across multiple files (e.g. into daily snapshots), which may be recovered by unioning tables stemming from the same repository. A second restriction limits the resulting search responses to 1000 files. This restriction makes the process of extracting a massive set of CSV files nontrivial, as detailed below.

First, we construct an initial ``topic query`` for each topic from WordNet restricted to files with the CSV format. For example, we retrieve CSV files that contain the word ``object'' by the query \texttt{q=``object'' extension:csv}. We exclude results from forked repositories to minimize table duplication. We execute this initial topic query through the GitHub Search API and get the initial response size of this query representing the number of GitHub URLs pointing to CSV files containing the word ``object''.

Since the API restricts the number of files per query to 1000 and many topic queries return around 100K files, we segment the initial queries. We use the ``size'' qualifier to perform this segmentation, and generate sequences of file size ranges (in bytes) proportional to the number of files in the initial response. This results in segmented topic queries like \texttt{q=``id'' extension:csv size:50..100}, \texttt{q=``id'' extension:csv size:100..150}, and so on. We execute all segmented queries and collect the paginated responses, each of which contains approximately 1000 URLs. We traverse the paginated responses to extract all URLs for a given topic. We then iteratively write the raw contents pointed to by the URLs to CSV files.

\subsection{Parsing and curating tables from CSV files} \label{sec:csv-parsing}

\paragraph{CSV parsing}
Once we have the CSV files, we parse them to tables using the CSV parser from the Pandas library, a widely used data processing and analysis library for Python~\cite{mckinney2011pandas}. We leverage the integrated functionality of Python's Sniffer tool to determine the delimiter of the CSV files. We parse tables assuming that the first rows correspond to the header rows, as is conventional for CSV files~\cite{tenneson2016csv}. Random samples of tables informed some exceptions to this; lines at the beginning of the file are skipped in case they are empty or start with a `\#`, which often indicates commented lines.

We remove rows in case they are considered ``bad lines'', such as empty lines, commented lines or lines with extra delimiters.
Additional experiments revealed that some tables include separation characters without any values at the end of all rows, resulting in a mismatch of the number of attributes and the number of values per row. We realigned the header and table values in these cases by removing the redundant separation characters. We discard CSV files that could not be parsed with the aforementioned rules. In total, our parsing procedure results in 99.3\% of the CSV files being parsed into tables.

\paragraph{Table filtering} We aim for \gittables to be a large-scale corpus of quality and relevant tables that can safely be used by the community. To this end, we first filter out tables from repositories without a license that allows the distribution of the repository contents. We find that roughly 16\% of the tables are associated with such a license, which is consistent with earlier studies. In our final corpus, we only publish tables coming from repositories with such a license.

We further curate the corpus by removing \textit{extremely small tables}~\cite{cafarella2008uncovering}, i.e. tables with less than two rows or two columns as these tables likely do not carry relevant data or were observed to contain singular text columns. We also remove tables if more than half of the column names are unspecified, or if any of the column names are not of the type \code{string}~\cite{mitlohner2016characteristics}. Lastly, to avoid inclusion of offensive content from social media platforms potentially stored in tables, as shown to be present in large collections of text extracted from  web pages~\cite{bender2021dangers}, we exclude tables of which a column name contains ``twitter'', ``tweet'', ``reddit'' or ``facebook''. Altogether, this curation pipeline filters out  9\% of the tables.

\paragraph{Content curation} Besides offensive content from social media, tables might contain personal data that is undesirable to disseminate beyond GitHub. To minimize the spread of personal identifiable information (PII), we anonymize tables that likely contain PII data informed by the semantic types from Schema.org. We do so, by replacing column values annotated with any of the PII semantic types (annotated as described in Section~\ref{sec:annotating-columns}) with fake values using the Faker library \cite{faker} as in Table \ref{tab:pii}. In case a column was annotated with the type \texttt{name}, we anonymize it only if it co-occurs with another PII semantic type as \texttt{name} does not necessarily indicate a person's name. In total 0.3\% of the columns in \gittables contain fake values. Given the relatively limited number of affected columns, this anonymization procedure does not significantly change the underlying data distribution of \gittables.

\begin{table}[!]
    \centering
    \caption{Semantic types associated with PII, percentage of columns annotated with each type (based on a subset), and Faker class used to generate fake column values.}
    \label{tab:pii}
    \begin{tabular}{lcl}
        \toprule
        \textbf{Semantic type} & \textbf{Percentage columns} & \textbf{Faker class} \\
        \midrule
        \code{name}            & 2.202\%     & \code{faker.name}\\
        \code{address}         & 0.163\%   & \code{faker.address}\\
        \code{person}          & 0.068\%  & \code{faker.name}\\
        \code{email }          & 0.042\%   & \code{faker.email}\\
        \code{birth date }     & 0.017\%   & \code{faker.date}\\
        \code{home location}   & 0.008\%   & \code{faker.city}\\
        \code{birth place}     &  0.003\%  & \code{faker.postcode}\\
        \code{postal code}     &  0.003\%  & \code{faker.city}\\
        \bottomrule
    \end{tabular}
\end{table}

\subsection{Annotating tables with column semantics}\label{sec:annotating-columns}
We annotate table columns in \gittables with semantic types extracted from DBpedia and Schema.org to facilitate its use-cases in applications like data preparation as explained in Section~\ref{sec:intro}. We provide rich metadata like hierarchical type relations, domains, and descriptions, if available, as well. This information can be exploited for training and evaluating models.



\paragraph{Semantic types} In total, we extracted 2831 properties from DBpedia that we use as semantic types. From Schema.org we included properties as well as types that together sum to 2637 semantic types. We provide annotations from both ontologies, a column may therefore be annotated with a semantic from DBpedia and one from Schema.org. 

Most semantic types from DBpedia relate to domains like \code{Person}, \code{Place} or \code{PopulatedPlace} while types in Schema.org are more scattered across domains topped by \code{CreativeWork}, \code{Organization}, \code{Person}, and \code{Offer}. Along with the semantic types, we attach their metadata like hierarchical relations and descriptions which can be exploited in model evaluation and training. For example, one could adopt a loss or evaluation function for a semantic type prediction model that favors a less granular type (e.g. the type \code{place} for a ground-truth column of type \code{city}), instead of predicting an unrelated type (e.g. \code{size}). We provide the following metadata per semantic type if available:

\begin{enumerate}[leftmargin=0.7cm, rightmargin=0em, itemsep=0.2em]
    \item semantic column type in English, e.g. \code{id} and \code{name},
    \item atomic type, e.g. \code{Number} and \code{Text},
    \item domain, e.g. \code{address} has domain \code{Person} and \code{Organization},
    \item superclass or superproperty, e.g. \code{product id} $\rightarrow$ \code{id}, and
    \item description, e.g. for \code{id}: ``The identifier property represents any kind of identifier for any kind of Thing, such as ISBNs, GTIN codes, UUIDs.''
\end{enumerate}

\paragraph{Annotation} Informed by analyses of public CSV files~\cite{mitlohner2016characteristics}, we preprocess the semantic types and table headers by replacing underscores and hyphens, splitting camel-cased combined words, and converting strings to lower case. Experiments showed that few column names with numbers were annotated with semantic types that coincidentally contain a number. For this reason, the annotation pipeline does not annotate column names that include numbers.

The original column names are useful indicators of what a column's data consists of and provide a proxy for human annotations given the involvement of humans in naming table columns. To provide relatively strict annotations, we therefore leverage the preprocessed column names directly and syntactically match them to semantic types in the ontologies~\cite{hulsebos2019sherlock}. We call this the syntactic annotation method.

Recent successes in language models create the opportunity to annotate columns taking semantics into account. We embed column names and semantic types using FastText~\cite{bojanowski2017enriching}, and match them to each other. We use the character-level n-gram FastText model pretrained on the Common Crawl corpus and take the match based on the highest cosine similarity

Although users can decide on a similarity threshold relevant to their tasks, we discard annotations with very low similarity scores to ensure that the annotations in \gittables are useful out of the box. We call this method the semantic annotation method.

%% file: 04_analysis.tex
\section{Analysis}\label{sec:analysis}
Here, we analyse the first version of \gittables of 1M tables over 97 topic subsets, among the larger sets: ``thing'', ``object'', and ``id''. Once complete, \gittables will consist of approximately 10M tables. Tables analysed in this paper are distributed as a separate version: \gittables 1M.

\subsection{Corpus statistics}

\begin{figure}
    \centering
    \begin{subfigure}{0.44\columnwidth}
        \includegraphics[width=1\columnwidth]{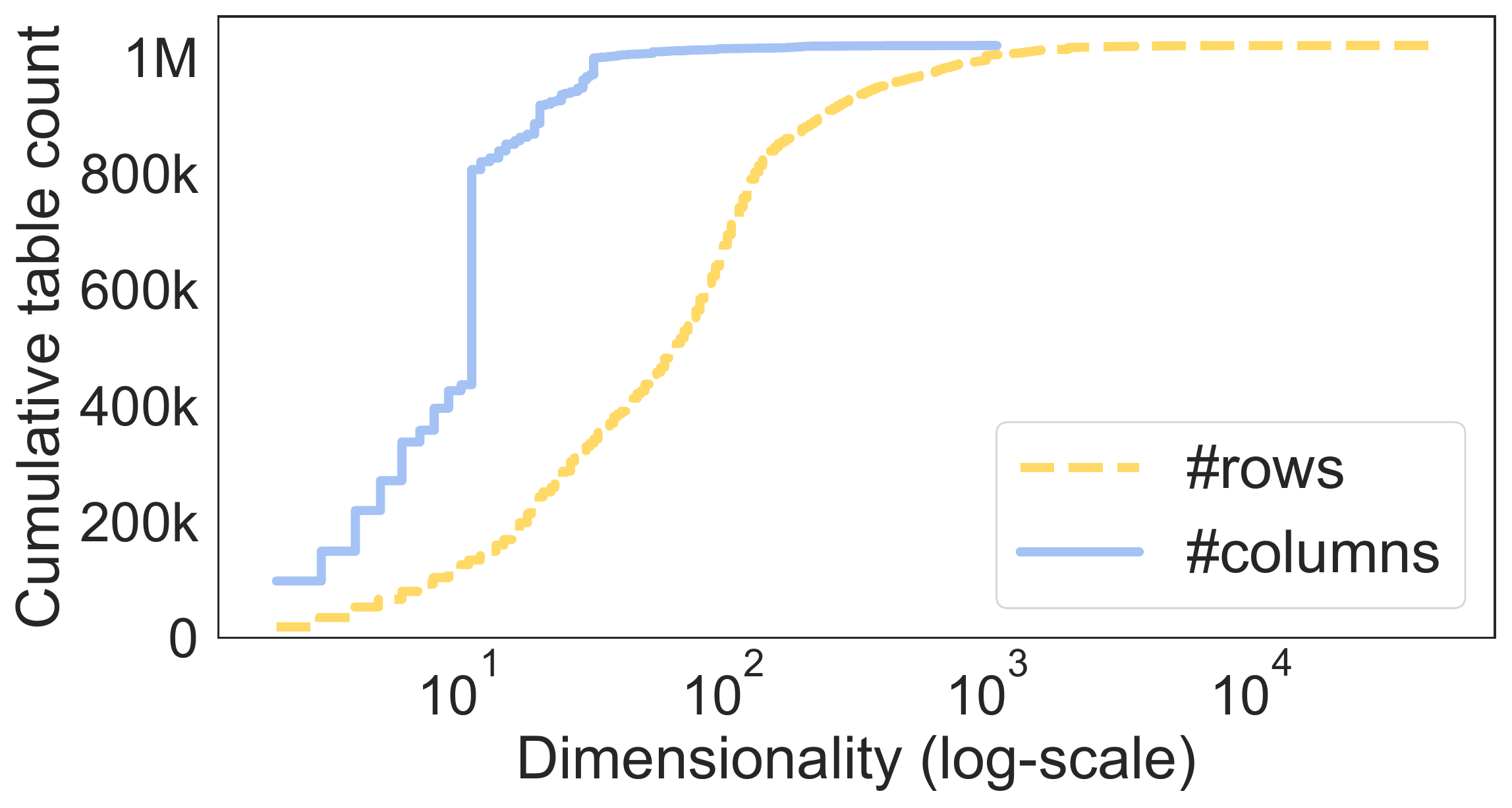}
        \caption{}
        \label{fig:dimension-distributions}
    \end{subfigure}
    \begin{subfigure}{0.26\columnwidth}
        \centering
      \includegraphics[width=1\columnwidth]{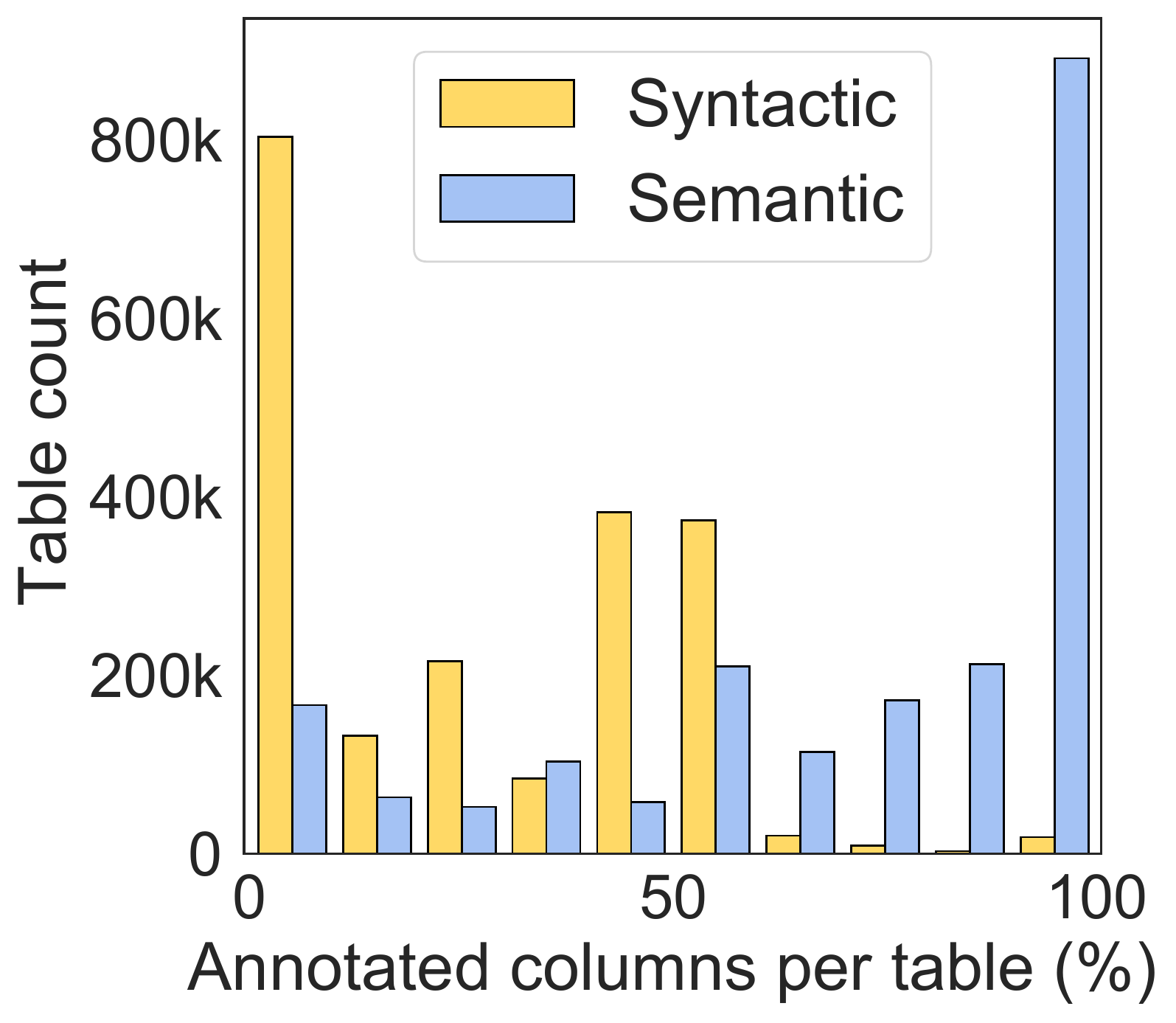}
      \caption{}
      \label{fig:percentage-annotated-cols} 
    \end{subfigure}\hfill%
    \begin{subfigure}{0.26\columnwidth}
        \centering
      \includegraphics[width=1\columnwidth]{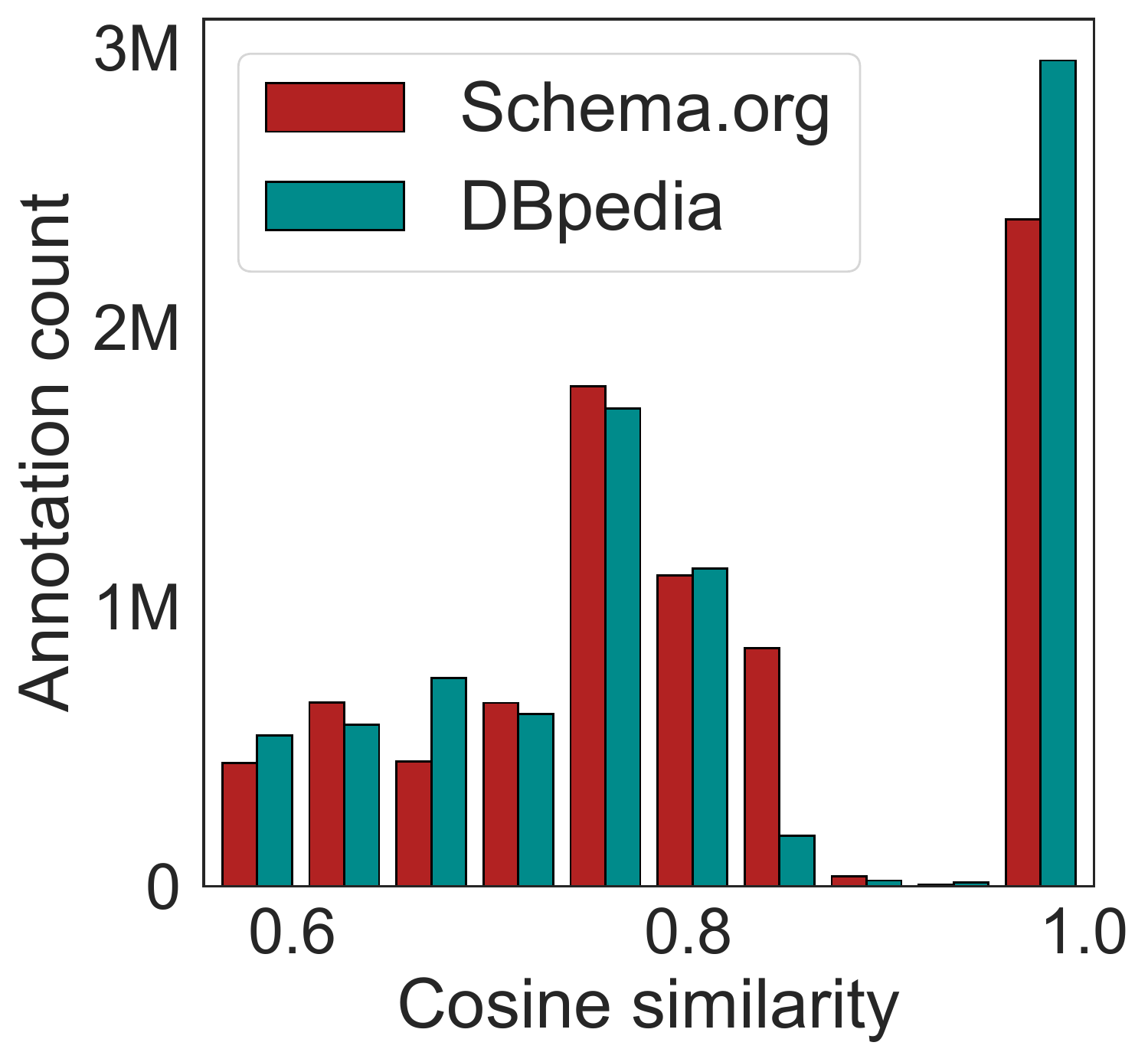}
      \caption{}
      \label{fig:semantic-similarities}
    \end{subfigure}
    \caption{(a) Cumulative table counts across table dimensions, illustrating the spread around the  mean column count (12) and mean row count (142). (b) Percentage annotated columns per table, for each annotation method. The semantic method yields on average more annotations per table. (c) Cosine similarity of annotated columns (both ontologies). The peak at 1 relates to high syntactic resemblance of column names and types.}
\end{figure}

\paragraph{Table statistics}
The total analysis set of 1,021,143 tables consists of 144,833,144 rows and 12,369,120 columns, on average 142 rows and 12 columns (see Tables~\ref{tab:table-corpora} and ~\ref{tab:annotated-table-corpora} for comparisons across corpora). On average, tables consist of 1,038 cells. Figure~\ref{fig:dimension-distributions} shows that the distribution of table dimensions has long tails over the number of rows and the number of columns. Overall these dimensions differ significantly from Web table corpora, which have 5 columns and 15 rows on average, and come closer to dimensions of typical database tables. As summarized in Table~\ref{tab:data-types}, almost 58\% of the columns are inferred to have numeric values versus 41\% textual, which differs from the roughly 50\%-50\% distribution in Web tables. Together, these statistics contribute to criteria C1 and C2.

\begin{table}
    \centering
    \caption{Distribution of atomic data types showing the difference in numeric versus string data between \gittables and WDC WebTables.}
    \label{tab:data-types}
    \vspace{-0.1cm}
    \begin{tabular}{l c c}
        \toprule
        \textbf{Atomic data type} & \textbf{GitTables} & \textbf{WDC WebTables} \\
        \midrule
        Numeric & 57.9\% & 51.4\% \\
        String & 41.6\% & 47.4\% \\
        Other & 0.5\% & 1.2\% \\
        \bottomrule
    \end{tabular}
\end{table}

The majority of the tables originate from distinct repositories as 75\% of the source repositories contribute at most 5 tables. With an overall average of 34 tables per repository, a few repositories contribute a large number of tables in this first version of \gittables. Manual inspection revealed that such repositories contain snapshots of the same or similar databases\footnote{For example, the repository: \url{ https://github.com/Emanuele-Falzone-PhD-Thesis/srbench}}. These tables, and the corresponding source URL referring to the associated repository, can be used for constructing larger tables through unions and joins.

We observed the advantage of separating the tables from the query topics. For example, the query for ``organism'' tables retrieves many tables related to biological and medical entities, of which a typical one is shown in Figure~\ref{fig:annotation-example}. Such subsets can be leveraged for training domain-specific models, or to incorporate these topics as semantics for table embeddings.

\paragraph{Annotation statistics} The syntactic method annotates, on average across the two ontologies, 730K out of 1M tables with at least 1 column annotation. On average, this yields 2.7M annotations with 756 unique types. The semantic annotation approach annotated on average 960K tables, yielding a total of 8.4M column annotations across 2.4K unique types on average. Statistics with regard to the different ontologies (DBpedia and Schema.org) are presented in Table \ref{tab:annotation-statistics}. Depending on the use case, users can select a suitable ontology and annotation method to filter the relevant annotations for each table.

\begin{table}
    \centering
    \caption{\small Statistics of annotations by method and ontology.}
    \label{tab:annotation-statistics}
    \begin{tabular}{l c c c c}
    \toprule
       \multirow{2}{*}{\textbf{Statistic}} & \multicolumn{2}{c}{\textbf{Syntactic}} & \multicolumn{2}{c}{\textbf{Semantic}} \\
       & \textbf{DBpedia} & \textbf{schema.org} & \textbf{DBpedia} & \textbf{schema.org} \\
        \midrule
       \# annotated tables & 723K & 738K & 958K & 962K \\
       \# annotated columns & 2.9M & 2.4M  & 8.5M & 8.4M \\
       \# types & 835 & 677 & 2.4K & 2.4K \\
       \# types (\#columns $>$ 1K) & 96 & 83 & 432 & 491 \\
    \bottomrule
    \end{tabular}
    \vspace{.3cm}
\end{table}

If column context is of importance, for example for contextual models, a high table-annotation coverage is key. We find that the semantic method yields annotations for, on average, 71\% of the table columns, while the syntactic method annotates 26\%. Figure~\ref{fig:percentage-annotated-cols} shows the overall distribution of the percentage of annotated columns per table aggregated over both ontologies.

The cosine-similarity scores, that we attach with all semantic annotations, reflect the confidence of each annotation. From the distribution of these similarity scores as shown in Figure~\ref{fig:semantic-similarities}, we observe that many annotations have a similarity score around 1, which indicates syntactic resemblance, while the remaining distribution is centered around 0.75. Users \gittables can set the desired threshold based on their needs and, for example, select only annotations with a high similarity score reflecting annotations with high confidence.

\subsection{Corpus content}\label{sec:contant-topical-coverage}

\paragraph {Table content and topical coverage}
Beyond the structural properties of tables in \gittables, we compare their contents to tables from VizNet which combines most existing corpora~\cite{hu2019viznet}. We interpret the comparison of table contents as a data shift detection problem and evaluate if the data distributions significantly differ by training a domain classifier~\cite{rabanser2019failing}. For this, we randomly sample 5K deduplicated columns from each corpus and extract 1,188 features as used for training semantic column type detection model Sherlock~\cite{hulsebos2019sherlock}, capturing column-level statistics like column entropy and skewness, aggregations from word-embeddings, and aggregated character-level statistics (e.g. number of `@' characters per cell).

We then train a Random Forest classifier with default settings to separate whether a column originated from VizNet or from \gittables. Using a 10-Fold cross-validation setup, we find that this domain classifier is able to predict for 93\% ($\pm$0.04) of the columns from which corpus it originates. This separability indicates the different data distributions in \gittables and VizNet, hence the complementary value of \gittables.

\begin{figure}
    \centering
    \includegraphics[width=0.95\columnwidth]{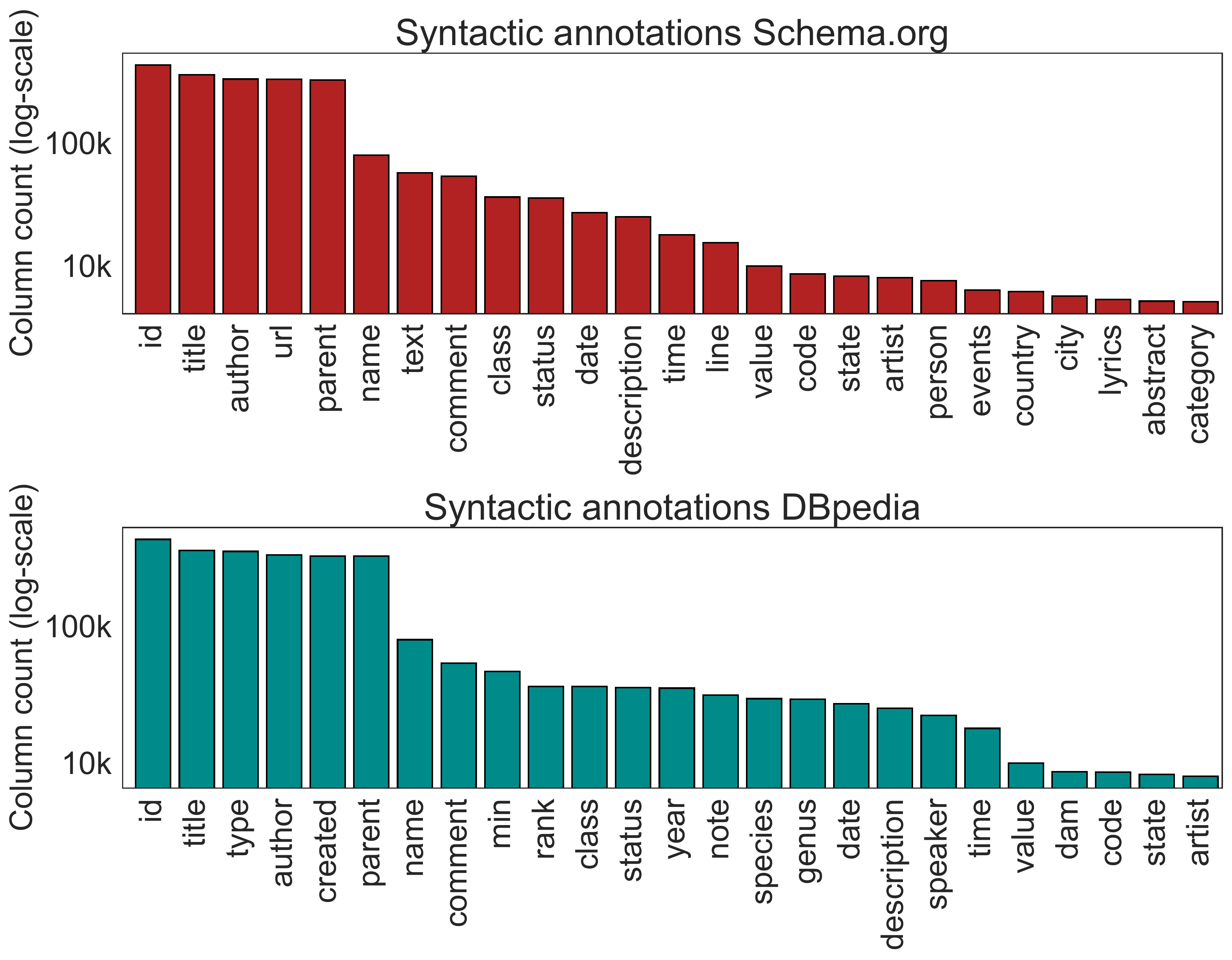}
    \caption{Column annotation counts of top 25 column semantic types across the two annotation methods and two target ontologies, illustrating the semantic coverage and diversity of \gittables.}
    \label{fig:type-distributions}
\end{figure}

The columns in the WDC WebTables 2012 corpus have also been matched to DBpedia which resulted in the top 10 semantic types: \code{name}, \code{date}, \code{title}, \code{artist}, \code{description}, \code{size}, \code{type}, \code{location}, \code{model}, and \code{year}. For \gittables, the topical coverage and diversity of semantic types per annotation method and ontology is illustrated in Figure~\ref{fig:type-distributions}. Although top types in both corpora overlap such as \code{name} and \code{title}, we observe clear differences with top types in \gittables like \code{id}, \code{type}, \code{value}, and \code{min}. Especially the dominant type \code{id}, one of the most common types in typical databases, indicates that \gittables meets criteria C2 and C3.

\paragraph{Content biases} The analysis of the topical distribution present in \gittables shows a few semantic types that could potentially make methods informed by \gittables lean towards certain subpopulations, industries, or geographic areas. To further profile our dataset to this end, we adopted 2 categories proposed in~\cite{wang2020revise}, being ``person'' and ``geography'' and analyse a subset of tables on biases towards these categories.

In Table \ref{tab:bias}, we give a deeper understanding of distributions along these dimensions informed by column values associated with relevant semantic types from Schema.org. This analysis confirms that tables in \gittables containing geographic data (slightly over 1\%), primarily represent English-speaking countries and cities. The value distribution of the few columns in \gittables with semantic types that reflect population segments, e.g. \code{gender}, \code{race}, and \code{ethnicity}, indicate a relatively higher concentration of data representing Western countries (see Table~\ref{tab:bias}).

\begin{table*}
    \centering
    \caption{Semantic types indicating subregions and subpopulations along with the most frequent column values. Tables with these types primarily represent Western and English-speaking regions and populations.}
    \label{tab:bias}
    \begin{threeparttable}
    \begin{tabular}{lcl}
    \toprule
    \textbf{Semantic type} & \textbf{Percentage columns} & \textbf{Frequent values (most frequent first)} \\
    \midrule
    \code{country} & 0.086\% & United States\tnote{2}, Canada, Belgium, Germany\\
    \code{city} & 0.056\% & New York (X), London, Coquitlam, Cambridge\\
    \code{gender} & 0.040\% & Male, Female, F, M\\
    \code{ethnicity} & 0.030\% & French, Dutch, Spanish, Mexican\\
    \code{race} & 0.007\% & Men, Human, White\\
    \code{nationality} & 0.003\% & Hispanic, White, Caucasian (White)\\
    \bottomrule
    \end{tabular}
    \begin{tablenotes}
        \item[2]``United States'' counts are merged with counts of ``USA''.
    \end{tablenotes}
    \end{threeparttable}
    \vspace{-.3cm}
\end{table*}

\subsection{Annotation quality}

As we have no access to the ground truth of semantic column types, we use the T2Dv2 benchmark hand-labeled with DBpedia types to evaluate our annotation approaches~\cite{t2dv2}. Although the annotation quality on T2Dv2 does not ensure the quality of annotations of \gittables, it is a good proxy for the annotation quality produced by our approaches. We consider table columns from files that we could parse and that were annotated by T2Dv2 as well as our annotation approaches. In total, we have 321 and 187 columns for evaluation with the semantic and syntactic approaches, respectively.

We find that our semantic approach yields in 54\% (173 columns) the same annotation as T2Dv2. From the incorrect annotations, our approach annotated 47\% (69 columns) with a DBpedia type that syntactically matches the column name as the corresponding similarity scores are 1.0. For example, our semantic approach annotated a column with cities (e.g. Pittsburgh, Buffalo) named ``City'' with the type \code{city} while T2Dv2 annotated this column with the less granular type \code{location}. This motivated a manual review of the 148 columns for which we find different annotations between the semantic approach and the annotations from T2Dv2\footnote{The reviews can be found on \url{ https://gittables.github.io}.}. Based on our manual review (n=3), we find that on average in 63 ($\pm$ 14) out of 148 the semantic approach yields better annotations, in 37 ($\pm$ 3) out of 148 the T2Dv2 annotations were clearly better. In 33 columns ($\pm$ 14) they were just as good or bad, and undetermined in 15 cases ($\pm$ 17).

The syntactic approach yields in 61\% (114 columns) the same annotation as T2Dv2.
At first, we found occurrences where the syntactic annotations were better. For example, a column with Latin names of birds named ``Latin name'' was annotated as \code{synonym} instead of DBpedia type \code{latin name}. We therefore followed the same manual review process as for the semantic approach to review 73 annotations where our pipeline yields annotations different from T2Dv2. Based on inter-annotator agreement we find that the syntactic approach clearly yields more accurate annotations for 21 columns and T2Dv2 had better annotations for 9 columns.

These findings indicate that our annotation approaches provide the necessary quality for training and evaluating DL models (C4). We also observe that the annotations of the T2Dv2 benchmark might require a review and potentially revision in future work.

%% file: 05_applications.tex
\section{Applications}\label{sec:applications}

In this section, we demonstrate how \gittables can be used for models for semantic column type detection and schema completion, and discuss its use for data management benchmarks in enterprise contexts. We note that the utility of \gittables stretches beyond these use cases to a wide range of data management tasks.

\subsection{Semantic column type detection}
Capturing the semantics of a table through its semantic column types is instrumental for many analytical tasks like data exploration as discussed in Section~\ref{sec:intro}, but also for GDPR compliant data processing like anonymizing PII data as implemented for \gittables. Many enterprise databases however lack descriptive or standardized column names across tables.
Column- and table representations learned from large table corpora have shown to accurately represent the semantics of Web tables~\cite{wang2021tcn}, and we believe that these models easily extend to domains beyond the Web when trained on \gittables.

To detect the semantic types of columns in a given table, we train Sherlock~\cite{hulsebos2019sherlock}, a deep learning model for semantic column type detection, on table columns from \gittables. For this experiment, we select five semantic types \code{address}, \code{class}, \code{status}, \code{name}, and \code{description} and randomly sampled 500 deduplicated columns per type from our analysis subset. For each column we extract the same low-level column features as used in Section~\ref{sec:contant-topical-coverage}, i.e. paragraph embeddings, word embeddings, character level counts, and aggregated statistics. We train Sherlock using a 5-Fold cross-validation setup. As Table \ref{tab:semantic-type-detection} shows, this model achieves a macro F1 score of on average 0.86 ($\pm$ 0.02) illustrating the utility of training a model on \gittables for semantic column type detection.

We repeat this experiment with columns from VizNet~\cite{hu2019viznet}, which combines tables from all existing table corpora. To ensure a fair comparison, we sample 500 deduplicated columns per type of the exact same types. This model yields a macro F1 score of 0.77 ($\pm$ 0.02) on samples from VizNet. However, when evaluating the VizNet-trained model on columns from \gittables, we find that it does not generalize well to samples from \gittables as the macro F1 score drops to 0.66 ($\pm$ 0.03). This gap confirms that models trained on Web tables alone, may not generalize beyond the Web.

\begin{table}[!]
\centering
\caption{F1 scores of semantic type detection models trained and evaluated on different corpora. These results suggest that models trained on GitTables can achieve high predictive performance, but that models trained on Web tables may not generalize to \gittables.}
  \label{tab:semantic-type-detection}
  \begin{tabular}{llc}
    \toprule
    \textbf{Train corpus}  & \textbf{Evaluation corpus} & \textbf{F1-score (macro)} \\
    \midrule
    GitTables  & GitTables & 0.86  \\
    VizNet     & VizNet    & 0.77  \\
    VizNet     & GitTables & 0.66 \\
    \bottomrule
  \end{tabular}
  \vspace{-0.3cm}
\end{table}

\subsection{Schema completion}\label{sec:schema-autocompletion}

Inspired by Cafarella et al.~\cite{cafarella2008webtables}, we use \gittables to inform completions of schema prefixes. Schema completion empowers many applications among which are general data- and knowledge base design, and data augmentation. We implement NearestCompletion as in Algorithm \ref{alg:schema-completion}. Our algorithm completes a given (target) schema prefix of length $N$ based on its similarity with schema prefixes from \gittables. We embed attributes from the target schema prefix of length $N$ and schemas from \gittables using the Universal Sentence Encoder (USE)~\cite{cer-etal-2018-universal} which works well for multi-word attributes. Similar to how search engines provide multiple suggestions for search queries, we return the $k$ ``nearest completions''. The nearest completions are based on schemas in \gittables with the smallest average cosine distance between the first $N$ attributes of these schemas and attributes of the target prefix of length $N$.

To evaluate the utility of \gittables for schema completion in typical industry contexts, we use schema prefixes from three actual table schemas from the CTU Prague Relational Learning Repository~\cite{motl2015ctu}\footnote{The ``employees'' table of the Employee database, the ``orders'' table from the ClassicalModels database, and the ``WorkOrder'' table from the AdventureWorks database.}. We extract prefixes of length $N=3$ and return the $k=10$ nearest completions as commonly used in information retrieval systems. The relevance of the suggested completions is calculated as the highest cosine similarity between USE embeddings of the original full schemas of the target prefix and the full schemas of the $10$ suggested completions from \gittables. A few attributes of the most relevant completion are presented in the second column of Table~\ref{tab:schema-completion}.

\begin{table*}
\centering
\caption{
Suggested completions from \gittables (second column) for target schema prefixes from the CTU database, along with the cosine similarity between the original full schema of the target prefix and the full schema of the suggested completion. The completions are relevant within the context of the schema prefix as reflected by the cosine similarities around 0.5 on a range of [-1,1].}
  \label{tab:schema-completion}
  \footnotesize
  \begin{tabular}{llc}
    \toprule
    \textbf{Header prefix}  & \textbf{Attributes from the nearest completion} & \textbf{Cosine similarity} \\
    \midrule
    emp\_no, birth\_date, first\_name  &
    Title, TitleOfCourtesy, Address, HireDate, City & 0.44 \\
    orderNumber, orderDate, requiredDate  &  ORDER\_TRACKING\_NUMBER, ORDER\_TOTAL & 0.50 \\
    WorkOrderID, ProductID, OrderQty & productType, inventoryId, articleId, productName & 0.53\\
    \bottomrule
    \normalsize
  \end{tabular}
  \vspace{-.4cm}
\end{table*}

\begin{algorithm}
    
    \KwData{Schema prefix $p$, set of schemas $\mathcal{S}$ from \gittables, number of completions $k$}
    \KwResult{Schema completions $c$}
    
    $N$ = |$p$|
    
    set of schema completions $\mathcal{C}$ = $\emptyset$
    
    embedded schema prefix $p_e$ = [embed($a$) for attribute $a$ $\in p$]
    
    \For{schema $s \in \mathcal{S}$}{
        embedded schema $s_e$ = [embed($a$) for attribute $a$ $\in s$]
        
        average prefix distance $d$ = $\frac{\sum\limits^{N}_{i=1}\text{cosine distance}(p_e[i], s_e[i])}{N}$
        
        $\mathcal{C}$.append($s$, $d$)
    }
    
    sort $\mathcal{C}$ in ascending order of $d$
    
    $c$ = $\mathcal{C}$.first($k$)
    
    \Return{$c$}
    
    \caption{\textbf{Schema completion procedure for returning $k$ most relevant completions for an input schema prefix based on the cosine similarity with schema prefixes in \gittables.}\label{alg:schema-completion}}
    
\end{algorithm}

Table \ref{tab:schema-completion} presents the target prefixes of the considered tables, attributes from the suggested completions, and the cosine similarities of the entire schemas. The semantics of these suggestions are clearly relevant to the target prefixes. For example, attributes from the most similar schema completion for the prefix from the ``employees'' table clearly relate to personal employee details such as ``HireDate'' and ``Title''. This relevance is also reflected by the cosine similarities between the entire original schemas and suggested schemas for the completions which, on average, is 0.49 on a scale of [-1,1]. These results underline the utility of \gittables as a resource for schema completion methods in typical databases.

\subsection{Data management benchmarks}
\vspace{-0.1cm}

\gittables provides a rich source for database-like tables that can be used to compile subsets for evaluating data management tasks such as table understanding and data search.

\begin{figure}
    \centering
    \vspace{-0.1cm}
    \begin{subfigure}{0.4\columnwidth}
    \includegraphics[width=1\columnwidth]{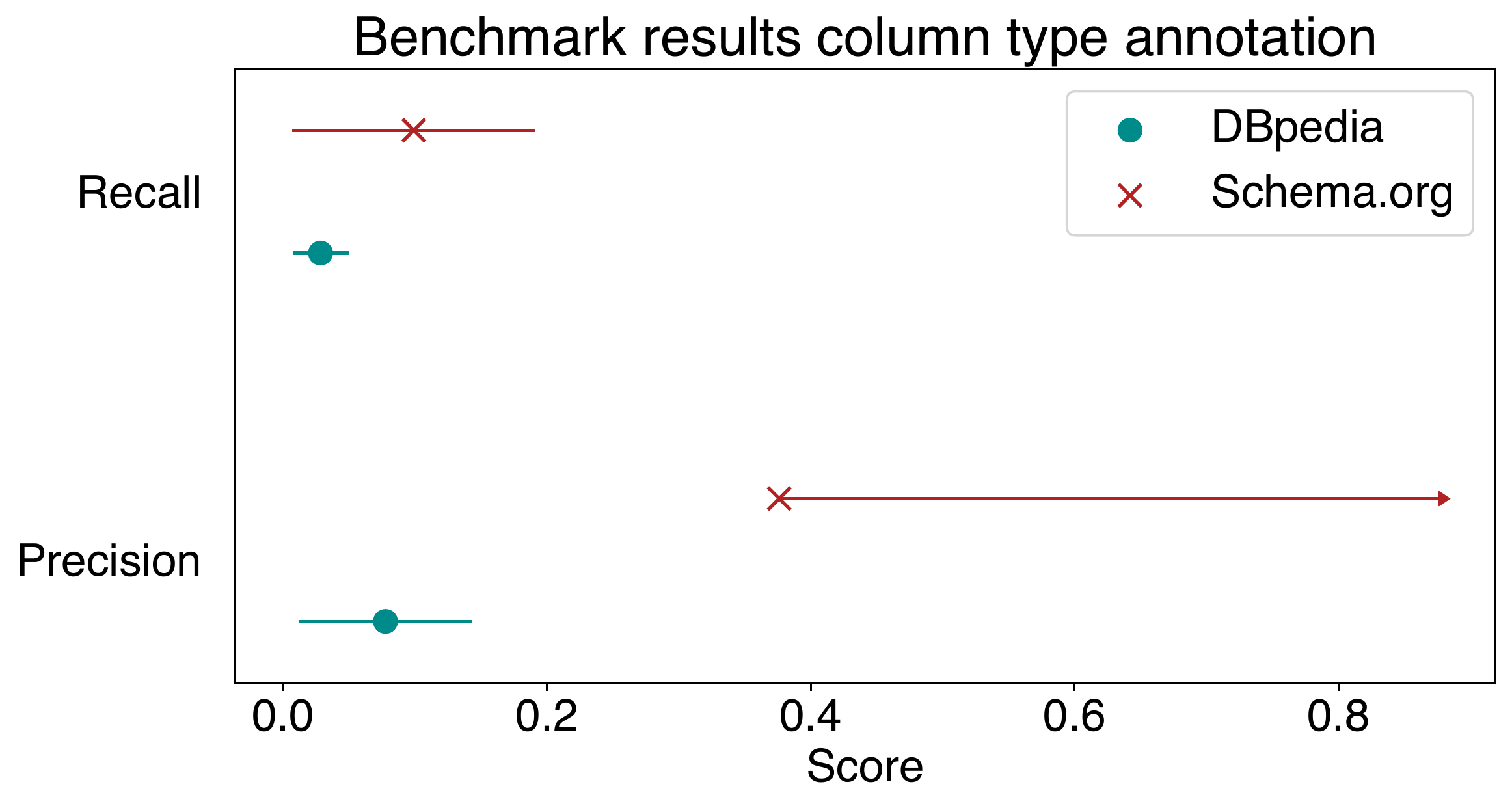}
    \vspace{-0.6cm}
    \caption{}
    \label{fig:applications-semtab}
    \end{subfigure}\hfill
    \begin{subfigure}{0.6\columnwidth}
    \includegraphics[width=1\columnwidth]{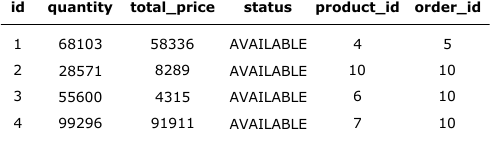}
    \vspace{-0.6cm}
    \caption{}
    \label{fig:nl-data-search}
    \end{subfigure}
    \vspace{-0.35cm}
    \caption{(a) Results from the SemTab '21 table-to-KG matching challenge for \gittables. The low precision and recall highlight the need for systems to better match table columns to KGs even if table values are not easily linked to KGs, as is the case for \gittables. (b) A table retrieved for search query ``\textit{status and sales amount per product}'' indicating the relevance of \gittables for benchmarking data search systems in enterprise contexts.}
    \vspace{-0.45cm}
\end{figure}

We created a benchmark dataset for semantic column type detection. We manually curated 1,101 tables, each having at least 3 columns and 5 rows. The target columns are associated with 122 distinct types from DBpedia and 59 types from Schema.org, as obtained with the syntactic annotation method. We envision that this benchmark dataset will stimulate the development of systems for enhancing knowledge graphs from novel data sources~\cite{weikum2021knowledge}. A first step in this direction is the inclusion of this dataset in the SemTab challenge for table-to-KG matching~\cite{semtab2021cutrona}. The results on this dataset already reveal open challenges for this task, as many systems rely on matching column cell values to KG entities. This typically does not work for tables from \gittables, as reflected by the low scores across systems (4 systems for DBpedia, and 3 for Schema.org), shown in Figure~\ref{fig:applications-semtab}. The average precision on the Schema.org annotations is slightly higher due to pattern matching methods that detected few structural types well.
Overall, our benchmark dataset poses new challenges for systems to match tables from sources beyond the Web to KGs.

Beyond table-to-KG matching, \gittables provides the basis for benchmarking data management tasks like data search. Evaluating systems on their search performance with \gittables is critical for understanding if they generalize to e.g. enterprise databases. To illustrate the relevance of \gittables for benchmarking this task, we implement a search procedure similar to Algorithm~\ref{alg:schema-completion} but instead embed entire table schemas and compare them with embedded search queries in natural language. Figure~\ref{fig:nl-data-search} shows a typical database table with product order data which is retrieved for the query ``\textit{status and sales amount per product}''. To develop this benchmark dataset further, one could collect a set of tables and queries and rank the most relevant tables for each query.

%% file: 07_conclusion.tex
\section{Conclusion} \label{sec:conclusion}



Relational databases are the bread and butter of enterprise applications. In this paper we introduce \gittables, a  corpus of 1M relational tables complementing existing Web table corpora available to researchers. We extract tables from CSV files in open-source repositories on GitHub, curate the corpus for quality control, and enhance tables with semantic column type annotations from DBpedia and Schema.org.
We analyse the corpus statistics, table contents, and annotation quality, which illustrate the complementary value of \gittables with respect to existing table corpora. We show that \gittables can be leveraged for 1) learned models for semantic column type detection with complementary coverage compared to existing corpora, 2) accurately suggesting schema completions in typical database contexts, and 3) developing challenging benchmarks for data management tasks. In future work, we are keen on exploring the broader utility of GitTables, which stretches beyond these use cases to a wide range of data management tasks.


%% file: main.bbl

\begin{thebibliography}{49}


\ifx \showCODEN    \undefined \def \showCODEN     #1{\unskip}     \fi
\ifx \showDOI      \undefined \def \showDOI       #1{#1}\fi
\ifx \showISBNx    \undefined \def \showISBNx     #1{\unskip}     \fi
\ifx \showISBNxiii \undefined \def \showISBNxiii  #1{\unskip}     \fi
\ifx \showISSN     \undefined \def \showISSN      #1{\unskip}     \fi
\ifx \showLCCN     \undefined \def \showLCCN      #1{\unskip}     \fi
\ifx \shownote     \undefined \def \shownote      #1{#1}          \fi
\ifx \showarticletitle \undefined \def \showarticletitle #1{#1}   \fi
\ifx \showURL      \undefined \def \showURL       {\relax}        \fi
\providecommand\bibfield[2]{#2}
\providecommand\bibinfo[2]{#2}
\providecommand\natexlab[1]{#1}
\providecommand\showeprint[2][]{arXiv:#2}

\bibitem[\protect\citeauthoryear{Auer, Bizer, Kobilarov, Lehmann, Cyganiak, and
  Ives}{Auer et~al\mbox{.}}{2007}]%
        {dbpedia}
\bibfield{author}{\bibinfo{person}{S{\"o}ren Auer}, \bibinfo{person}{Christian
  Bizer}, \bibinfo{person}{Georgi Kobilarov}, \bibinfo{person}{Jens Lehmann},
  \bibinfo{person}{Richard Cyganiak}, {and} \bibinfo{person}{Zachary Ives}.}
  \bibinfo{year}{2007}\natexlab{}.
\newblock \showarticletitle{{DB}pedia{:} A nucleus for a web of open data}.
\newblock \bibinfo{journal}{\emph{{ISWC}}} (\bibinfo{year}{2007}),
  \bibinfo{pages}{722--735}.
\newblock


\bibitem[\protect\citeauthoryear{Bender, Gebru, McMillan-Major, and
  Shmitchell}{Bender et~al\mbox{.}}{2021}]%
        {bender2021dangers}
\bibfield{author}{\bibinfo{person}{Emily~M Bender}, \bibinfo{person}{Timnit
  Gebru}, \bibinfo{person}{Angelina McMillan-Major}, {and}
  \bibinfo{person}{Shmargaret Shmitchell}.} \bibinfo{year}{2021}\natexlab{}.
\newblock \showarticletitle{On the Dangers of Stochastic Parrots: Can Language
  Models Be Too Big?}. In \bibinfo{booktitle}{\emph{Proceedings of the 2021 ACM
  Conference on Fairness, Accountability, and Transparency}}.
  \bibinfo{pages}{610--623}.
\newblock


\bibitem[\protect\citeauthoryear{Bhagavatula, Noraset, and Downey}{Bhagavatula
  et~al\mbox{.}}{2013}]%
        {bhagavatula2013wikitables}
\bibfield{author}{\bibinfo{person}{Chandra~Sekhar Bhagavatula},
  \bibinfo{person}{Thanapon Noraset}, {and} \bibinfo{person}{Doug Downey}.}
  \bibinfo{year}{2013}\natexlab{}.
\newblock \showarticletitle{Methods for exploring and mining tables on
  wikipedia}. In \bibinfo{booktitle}{\emph{Proceedings of the ACM SIGKDD
  workshop on interactive data exploration and analytics}}.
  \bibinfo{pages}{18--26}.
\newblock


\bibitem[\protect\citeauthoryear{Birhane and Prabhu}{Birhane and
  Prabhu}{2021}]%
        {birhane2021large}
\bibfield{author}{\bibinfo{person}{Abeba Birhane} {and}
  \bibinfo{person}{Vinay~Uday Prabhu}.} \bibinfo{year}{2021}\natexlab{}.
\newblock \showarticletitle{Large image datasets: A pyrrhic win for computer
  vision?}. In \bibinfo{booktitle}{\emph{2021 IEEE Winter Conference on
  Applications of Computer Vision (WACV)}}. IEEE, \bibinfo{pages}{1536--1546}.
\newblock


\bibitem[\protect\citeauthoryear{Bojanowski, Grave, Joulin, and
  Mikolov}{Bojanowski et~al\mbox{.}}{2017}]%
        {bojanowski2017enriching}
\bibfield{author}{\bibinfo{person}{Piotr Bojanowski}, \bibinfo{person}{Edouard
  Grave}, \bibinfo{person}{Armand Joulin}, {and} \bibinfo{person}{Tomas
  Mikolov}.} \bibinfo{year}{2017}\natexlab{}.
\newblock \showarticletitle{Enriching word vectors with subword information}.
\newblock \bibinfo{journal}{\emph{Transactions of the Association for
  Computational Linguistics}}  \bibinfo{volume}{5} (\bibinfo{year}{2017}),
  \bibinfo{pages}{135--146}.
\newblock


\bibitem[\protect\citeauthoryear{Brown, Mann, Ryder, Subbiah, Kaplan, Dhariwal,
  Neelakantan, Shyam, Sastry, Askell, et~al\mbox{.}}{Brown
  et~al\mbox{.}}{2020}]%
        {brown2020gpt3}
\bibfield{author}{\bibinfo{person}{Tom~B Brown}, \bibinfo{person}{Benjamin
  Mann}, \bibinfo{person}{Nick Ryder}, \bibinfo{person}{Melanie Subbiah},
  \bibinfo{person}{Jared Kaplan}, \bibinfo{person}{Prafulla Dhariwal},
  \bibinfo{person}{Arvind Neelakantan}, \bibinfo{person}{Pranav Shyam},
  \bibinfo{person}{Girish Sastry}, \bibinfo{person}{Amanda Askell},
  {et~al\mbox{.}}} \bibinfo{year}{2020}\natexlab{}.
\newblock \showarticletitle{Language models are few-shot learners}.
\newblock \bibinfo{journal}{\emph{arXiv preprint arXiv:2005.14165}}
  (\bibinfo{year}{2020}).
\newblock


\bibitem[\protect\citeauthoryear{Cafarella, Halevy, Lee, Madhavan, Yu, Wang,
  and Wu}{Cafarella et~al\mbox{.}}{2018}]%
        {webtablestenyears}
\bibfield{author}{\bibinfo{person}{Michael Cafarella}, \bibinfo{person}{Alon
  Halevy}, \bibinfo{person}{Hongrae Lee}, \bibinfo{person}{Jayant Madhavan},
  \bibinfo{person}{Cong Yu}, \bibinfo{person}{Daisy~Zhe Wang}, {and}
  \bibinfo{person}{Eugene Wu}.} \bibinfo{year}{2018}\natexlab{}.
\newblock \showarticletitle{Ten years of webtables}.
\newblock \bibinfo{journal}{\emph{Proceedings of the VLDB Endowment}}
  \bibinfo{volume}{11}, \bibinfo{number}{12} (\bibinfo{year}{2018}),
  \bibinfo{pages}{2140--2149}.
\newblock


\bibitem[\protect\citeauthoryear{Cafarella, Halevy, Wang, Wu, and
  Zhang}{Cafarella et~al\mbox{.}}{2008a}]%
        {carafella2008webtables}
\bibfield{author}{\bibinfo{person}{Michael~J Cafarella}, \bibinfo{person}{Alon
  Halevy}, \bibinfo{person}{Daisy~Zhe Wang}, \bibinfo{person}{Eugene Wu}, {and}
  \bibinfo{person}{Yang Zhang}.} \bibinfo{year}{2008}\natexlab{a}.
\newblock \showarticletitle{Web{T}ables: exploring the power of tables on the
  web}.
\newblock \bibinfo{journal}{\emph{Proceedings of the VLDB Endowment}}
  \bibinfo{volume}{1}, \bibinfo{number}{1} (\bibinfo{year}{2008}),
  \bibinfo{pages}{538--549}.
\newblock


\bibitem[\protect\citeauthoryear{Cafarella, Halevy, Wang, Wu, and
  Zhang}{Cafarella et~al\mbox{.}}{2008b}]%
        {cafarella2008webtables}
\bibfield{author}{\bibinfo{person}{Michael~J. Cafarella}, \bibinfo{person}{Alon
  Halevy}, \bibinfo{person}{Daisy~Zhe Wang}, \bibinfo{person}{Eugene Wu}, {and}
  \bibinfo{person}{Yang Zhang}.} \bibinfo{year}{2008}\natexlab{b}.
\newblock \showarticletitle{WebTables: Exploring the Power of Tables on the
  Web}.
\newblock \bibinfo{journal}{\emph{{PVLDB}}} (\bibinfo{year}{2008}),
  \bibinfo{pages}{538--549}.
\newblock


\bibitem[\protect\citeauthoryear{Cafarella, Halevy, Zhang, Wang, and
  Wu}{Cafarella et~al\mbox{.}}{2008c}]%
        {cafarella2008uncovering}
\bibfield{author}{\bibinfo{person}{Michael~J Cafarella},
  \bibinfo{person}{Alon~Y Halevy}, \bibinfo{person}{Yang Zhang},
  \bibinfo{person}{Daisy~Zhe Wang}, {and} \bibinfo{person}{Eugene Wu}.}
  \bibinfo{year}{2008}\natexlab{c}.
\newblock \showarticletitle{Uncovering the Relational Web.}. In
  \bibinfo{booktitle}{\emph{WebDB}}. \bibinfo{pages}{1--6}.
\newblock


\bibitem[\protect\citeauthoryear{Cer, Yang, Kong, Hua, Limtiaco, St.~John,
  Constant, Guajardo-Cespedes, Yuan, Tar, Strope, and Kurzweil}{Cer
  et~al\mbox{.}}{2018}]%
        {cer-etal-2018-universal}
\bibfield{author}{\bibinfo{person}{Daniel Cer}, \bibinfo{person}{Yinfei Yang},
  \bibinfo{person}{Sheng-yi Kong}, \bibinfo{person}{Nan Hua},
  \bibinfo{person}{Nicole Limtiaco}, \bibinfo{person}{Rhomni St.~John},
  \bibinfo{person}{Noah Constant}, \bibinfo{person}{Mario Guajardo-Cespedes},
  \bibinfo{person}{Steve Yuan}, \bibinfo{person}{Chris Tar},
  \bibinfo{person}{Brian Strope}, {and} \bibinfo{person}{Ray Kurzweil}.}
  \bibinfo{year}{2018}\natexlab{}.
\newblock \showarticletitle{Universal {S}entence {E}ncoder for {E}nglish}. In
  \bibinfo{booktitle}{\emph{Proceedings of the 2018 Conference on Empirical
  Methods in Natural Language Processing: System Demonstrations}}.
  \bibinfo{publisher}{Association for Computational Linguistics},
  \bibinfo{address}{Brussels, Belgium}, \bibinfo{pages}{169--174}.
\newblock
\urldef\tempurl%
\url{https://doi.org/10.18653/v1/D18-2029}
\showDOI{\tempurl}


\bibitem[\protect\citeauthoryear{Cutrona, Bianchi, Jim{\'e}nez-Ruiz, and
  Palmonari}{Cutrona et~al\mbox{.}}{2020}]%
        {cutrona2020tough}
\bibfield{author}{\bibinfo{person}{Vincenzo Cutrona}, \bibinfo{person}{Federico
  Bianchi}, \bibinfo{person}{Ernesto Jim{\'e}nez-Ruiz}, {and}
  \bibinfo{person}{Matteo Palmonari}.} \bibinfo{year}{2020}\natexlab{}.
\newblock \showarticletitle{Tough {T}ables: {C}arefully evaluating entity
  linking for tabular data}. In \bibinfo{booktitle}{\emph{International
  Semantic Web Conference}}. Springer, \bibinfo{pages}{328--343}.
\newblock


\bibitem[\protect\citeauthoryear{Cutrona, Chen, Efthymiou, Hassanzadeh,
  Jim{\'{e}}nez{-}Ruiz, Sequeda, Srinivas, Abdelmageed, Hulsebos, Oliveira, and
  Pesquita}{Cutrona et~al\mbox{.}}{2021}]%
        {semtab2021cutrona}
\bibfield{author}{\bibinfo{person}{Vincenzo Cutrona}, \bibinfo{person}{Jiaoyan
  Chen}, \bibinfo{person}{Vasilis Efthymiou}, \bibinfo{person}{Oktie
  Hassanzadeh}, \bibinfo{person}{Ernesto Jim{\'{e}}nez{-}Ruiz},
  \bibinfo{person}{Juan Sequeda}, \bibinfo{person}{Kavitha Srinivas},
  \bibinfo{person}{Nora Abdelmageed}, \bibinfo{person}{Madelon Hulsebos},
  \bibinfo{person}{Daniela Oliveira}, {and} \bibinfo{person}{Catia Pesquita}.}
  \bibinfo{year}{2021}\natexlab{}.
\newblock \showarticletitle{Results of SemTab 2021}. In
  \bibinfo{booktitle}{\emph{Proceedings of the Semantic Web Challenge on
  Tabular Data to Knowledge Graph Matching co-located with the 20th
  International Semantic Web Conference {(ISWC 2021)}, Virtual conference,
  October 27, 2021}} \emph{(\bibinfo{series}{{CEUR} Workshop Proceedings},
  Vol.~\bibinfo{volume}{3103})}. \bibinfo{publisher}{CEUR-WS.org},
  \bibinfo{pages}{1--12}.
\newblock
\urldef\tempurl%
\url{http://ceur-ws.org/Vol-3103/paper0.pdf}
\showURL{%
\tempurl}


\bibitem[\protect\citeauthoryear{Deng, Dong, Socher, Li, Li, and Fei-Fei}{Deng
  et~al\mbox{.}}{2009}]%
        {imagenet}
\bibfield{author}{\bibinfo{person}{Jia Deng}, \bibinfo{person}{Wei Dong},
  \bibinfo{person}{Richard Socher}, \bibinfo{person}{Li-Jia Li},
  \bibinfo{person}{Kai Li}, {and} \bibinfo{person}{Li Fei-Fei}.}
  \bibinfo{year}{2009}\natexlab{}.
\newblock \showarticletitle{Image{N}et: A large-scale hierarchical image
  database}. In \bibinfo{booktitle}{\emph{2009 IEEE conference on computer
  vision and pattern recognition}}. Ieee, \bibinfo{pages}{248--255}.
\newblock


\bibitem[\protect\citeauthoryear{Deng, Sun, Lees, Wu, and Yu}{Deng
  et~al\mbox{.}}{2020}]%
        {turl}
\bibfield{author}{\bibinfo{person}{Xiang Deng}, \bibinfo{person}{Huan Sun},
  \bibinfo{person}{Alyssa Lees}, \bibinfo{person}{You Wu}, {and}
  \bibinfo{person}{Cong Yu}.} \bibinfo{year}{2020}\natexlab{}.
\newblock \showarticletitle{{TURL}: {T}able {U}nderstanding through
  {R}epresentation {L}earning}.
\newblock \bibinfo{journal}{\emph{Proceedings of the VLDB Endowment}}
  \bibinfo{volume}{14}, \bibinfo{number}{3} (\bibinfo{year}{2020}),
  \bibinfo{pages}{307--319}.
\newblock


\bibitem[\protect\citeauthoryear{Devlin, Chang, Lee, and Toutanova}{Devlin
  et~al\mbox{.}}{2019}]%
        {devlin2019bert}
\bibfield{author}{\bibinfo{person}{Jacob Devlin}, \bibinfo{person}{Ming-Wei
  Chang}, \bibinfo{person}{Kenton Lee}, {and} \bibinfo{person}{Kristina
  Toutanova}.} \bibinfo{year}{2019}\natexlab{}.
\newblock \showarticletitle{BERT: Pre-training of Deep Bidirectional
  Transformers for Language Understanding}. In
  \bibinfo{booktitle}{\emph{Proceedings of the 2019 Conference of the North
  American Chapter of the Association for Computational Linguistics: Human
  Language Technologies, Volume 1}}. \bibinfo{pages}{4171--4186}.
\newblock


\bibitem[\protect\citeauthoryear{Eberius, Braunschweig, Hentsch, Thiele,
  Ahmadov, and Lehner}{Eberius et~al\mbox{.}}{2015}]%
        {dresdenwebtables}
\bibfield{author}{\bibinfo{person}{Julian Eberius}, \bibinfo{person}{Katrin
  Braunschweig}, \bibinfo{person}{Markus Hentsch}, \bibinfo{person}{Maik
  Thiele}, \bibinfo{person}{Ahmad Ahmadov}, {and} \bibinfo{person}{Wolfgang
  Lehner}.} \bibinfo{year}{2015}\natexlab{}.
\newblock \showarticletitle{Building the dresden web table corpus: A
  classification approach}. In \bibinfo{booktitle}{\emph{2015 IEEE/ACM 2nd
  International Symposium on Big Data Computing (BDC)}}. IEEE,
  \bibinfo{pages}{41--50}.
\newblock


\bibitem[\protect\citeauthoryear{{F}araglia and {Other
  Contributors}}{{F}araglia and {Other Contributors}}{2014}]%
        {faker}
\bibfield{author}{\bibinfo{person}{{D}aniele {F}araglia} {and}
  \bibinfo{person}{{Other Contributors}}.} \bibinfo{year}{2014}\natexlab{}.
\newblock \bibinfo{title}{{Faker}}.
\newblock
\newblock
\urldef\tempurl%
\url{https://github.com/joke2k/faker}
\showURL{%
\tempurl}


\bibitem[\protect\citeauthoryear{Guha, Brickley, and Macbeth}{Guha
  et~al\mbox{.}}{2016}]%
        {guha2016schema}
\bibfield{author}{\bibinfo{person}{Ramanathan~V Guha}, \bibinfo{person}{Dan
  Brickley}, {and} \bibinfo{person}{Steve Macbeth}.}
  \bibinfo{year}{2016}\natexlab{}.
\newblock \showarticletitle{Schema. org: evolution of structured data on the
  web}.
\newblock \bibinfo{journal}{\emph{Commun. ACM}} \bibinfo{volume}{59},
  \bibinfo{number}{2} (\bibinfo{year}{2016}), \bibinfo{pages}{44--51}.
\newblock


\bibitem[\protect\citeauthoryear{Hu, Gaikwad, Bakker, Hulsebos, Zgraggen,
  Hidalgo, Kraska, Li, Satyanarayan, and Demiralp}{Hu et~al\mbox{.}}{2019}]%
        {hu2019viznet}
\bibfield{author}{\bibinfo{person}{Kevin Hu}, \bibinfo{person}{Neil Gaikwad},
  \bibinfo{person}{Michiel Bakker}, \bibinfo{person}{Madelon Hulsebos},
  \bibinfo{person}{Emanuel Zgraggen}, \bibinfo{person}{C\'{e}sar Hidalgo},
  \bibinfo{person}{Tim Kraska}, \bibinfo{person}{Guoliang Li},
  \bibinfo{person}{Arvind Satyanarayan}, {and}
  \bibinfo{person}{{\c{C}}a{\u{g}}atay Demiralp}.}
  \bibinfo{year}{2019}\natexlab{}.
\newblock \showarticletitle{Viz{N}et: {T}owards a large-scale visualization
  learning and benchmarking repository}. In \bibinfo{booktitle}{\emph{{CHI}}}.
  \bibinfo{publisher}{ACM}.
\newblock


\bibitem[\protect\citeauthoryear{Hulsebos, Hu, Bakker, Zgraggen, Satyanarayan,
  Kraska, Demiralp, and Hidalgo}{Hulsebos et~al\mbox{.}}{2019}]%
        {hulsebos2019sherlock}
\bibfield{author}{\bibinfo{person}{Madelon Hulsebos}, \bibinfo{person}{Kevin
  Hu}, \bibinfo{person}{Michiel Bakker}, \bibinfo{person}{Emanuel Zgraggen},
  \bibinfo{person}{Arvind Satyanarayan}, \bibinfo{person}{Tim Kraska},
  \bibinfo{person}{{\c{C}}agatay Demiralp}, {and} \bibinfo{person}{C{\'e}sar
  Hidalgo}.} \bibinfo{year}{2019}\natexlab{}.
\newblock \showarticletitle{Sherlock: A deep learning approach to semantic data
  type detection}. In \bibinfo{booktitle}{\emph{Proceedings of the 25th ACM
  SIGKDD International Conference on Knowledge Discovery \& Data Mining}}.
  \bibinfo{pages}{1500--1508}.
\newblock


\bibitem[\protect\citeauthoryear{Jimenez-Ruiz, Hassanzadeh, Efthymiou, Chen,
  Srinivas, and Cutrona}{Jimenez-Ruiz et~al\mbox{.}}{2020}]%
        {jimenez2020semtab}
\bibfield{author}{\bibinfo{person}{Ernesto Jimenez-Ruiz},
  \bibinfo{person}{Oktie Hassanzadeh}, \bibinfo{person}{Vasilis Efthymiou},
  \bibinfo{person}{Jiaoyan Chen}, \bibinfo{person}{Kavitha Srinivas}, {and}
  \bibinfo{person}{Vincenzo Cutrona}.} \bibinfo{year}{2020}\natexlab{}.
\newblock \showarticletitle{Results of {S}em{T}ab 2020}. In
  \bibinfo{booktitle}{\emph{CEUR Workshop Proceedings}},
  Vol.~\bibinfo{volume}{2775}. \bibinfo{pages}{1--8}.
\newblock


\bibitem[\protect\citeauthoryear{Koesten, Vougiouklis, Simperl, and
  Groth}{Koesten et~al\mbox{.}}{2020}]%
        {koesten2020dataset}
\bibfield{author}{\bibinfo{person}{Laura Koesten}, \bibinfo{person}{Pavlos
  Vougiouklis}, \bibinfo{person}{Elena Simperl}, {and} \bibinfo{person}{Paul
  Groth}.} \bibinfo{year}{2020}\natexlab{}.
\newblock \showarticletitle{Dataset Reuse: Toward Translating Principles to
  Practice}.
\newblock \bibinfo{journal}{\emph{Patterns}} (\bibinfo{year}{2020}),
  \bibinfo{pages}{100136}.
\newblock


\bibitem[\protect\citeauthoryear{Langenecker, Sturm, Schalles, and
  Binnig}{Langenecker et~al\mbox{.}}{2021}]%
        {langenecker2021towards}
\bibfield{author}{\bibinfo{person}{Sven Langenecker},
  \bibinfo{person}{Christoph Sturm}, \bibinfo{person}{Christian Schalles},
  {and} \bibinfo{person}{Carsten Binnig}.} \bibinfo{year}{2021}\natexlab{}.
\newblock \showarticletitle{Towards Learned Metadata Extraction for Data
  Lakes}.
\newblock \bibinfo{journal}{\emph{BTW 2021}} (\bibinfo{year}{2021}).
\newblock


\bibitem[\protect\citeauthoryear{LeCun, Bengio, and Hinton}{LeCun
  et~al\mbox{.}}{2015}]%
        {lecun2015deep}
\bibfield{author}{\bibinfo{person}{Yann LeCun}, \bibinfo{person}{Yoshua
  Bengio}, {and} \bibinfo{person}{Geoffrey Hinton}.}
  \bibinfo{year}{2015}\natexlab{}.
\newblock \showarticletitle{Deep learning}.
\newblock \bibinfo{journal}{\emph{nature}} \bibinfo{volume}{521},
  \bibinfo{number}{7553} (\bibinfo{year}{2015}), \bibinfo{pages}{436}.
\newblock


\bibitem[\protect\citeauthoryear{Lehmberg, Ritze, Meusel, and Bizer}{Lehmberg
  et~al\mbox{.}}{2016}]%
        {Lehmberg:2016:WebTableCorpus}
\bibfield{author}{\bibinfo{person}{Oliver Lehmberg}, \bibinfo{person}{Dominique
  Ritze}, \bibinfo{person}{Robert Meusel}, {and} \bibinfo{person}{Christian
  Bizer}.} \bibinfo{year}{2016}\natexlab{}.
\newblock \showarticletitle{A Large Public Corpus of Web Tables Containing Time
  and Context Metadata}. In \bibinfo{booktitle}{\emph{{WWW} Companion}}.
  \bibinfo{pages}{75--76}.
\newblock


\bibitem[\protect\citeauthoryear{Li, He, and Ganjam}{Li et~al\mbox{.}}{2017}]%
        {li2017discoveringenterprise}
\bibfield{author}{\bibinfo{person}{Keqian Li}, \bibinfo{person}{Yeye He}, {and}
  \bibinfo{person}{Kris Ganjam}.} \bibinfo{year}{2017}\natexlab{}.
\newblock \showarticletitle{Discovering enterprise concepts using spreadsheet
  tables}. In \bibinfo{booktitle}{\emph{Proceedings of the 23rd ACM SIGKDD
  International Conference on Knowledge Discovery and Data Mining}}.
  \bibinfo{pages}{1873--1882}.
\newblock


\bibitem[\protect\citeauthoryear{McKinney et~al\mbox{.}}{McKinney
  et~al\mbox{.}}{2011}]%
        {mckinney2011pandas}
\bibfield{author}{\bibinfo{person}{Wes McKinney} {et~al\mbox{.}}}
  \bibinfo{year}{2011}\natexlab{}.
\newblock \showarticletitle{pandas: a foundational {P}ython library for data
  analysis and statistics}.
\newblock \bibinfo{journal}{\emph{Python for High Performance and Scientific
  Computing}} \bibinfo{volume}{14}, \bibinfo{number}{9} (\bibinfo{year}{2011}),
  \bibinfo{pages}{1--9}.
\newblock


\bibitem[\protect\citeauthoryear{Mitl{\"o}hner, Neumaier, Umbrich, and
  Polleres}{Mitl{\"o}hner et~al\mbox{.}}{2016}]%
        {mitlohner2016characteristics}
\bibfield{author}{\bibinfo{person}{Johann Mitl{\"o}hner},
  \bibinfo{person}{Sebastian Neumaier}, \bibinfo{person}{J{\"u}rgen Umbrich},
  {and} \bibinfo{person}{Axel Polleres}.} \bibinfo{year}{2016}\natexlab{}.
\newblock \showarticletitle{{C}haracteristics of open data {CSV} files}. In
  \bibinfo{booktitle}{\emph{2016 2nd International Conference on Open and Big
  Data (OBD)}}. IEEE, \bibinfo{pages}{72--79}.
\newblock


\bibitem[\protect\citeauthoryear{Motl and Schulte}{Motl and Schulte}{2015}]%
        {motl2015ctu}
\bibfield{author}{\bibinfo{person}{Jan Motl} {and} \bibinfo{person}{Oliver
  Schulte}.} \bibinfo{year}{2015}\natexlab{}.
\newblock \showarticletitle{The CTU prague relational learning repository}.
\newblock \bibinfo{journal}{\emph{arXiv preprint arXiv:1511.03086}}
  (\bibinfo{year}{2015}).
\newblock


\bibitem[\protect\citeauthoryear{M{\"u}hleisen and Bizer}{M{\"u}hleisen and
  Bizer}{2012}]%
        {muhleisen2012web}
\bibfield{author}{\bibinfo{person}{Hannes M{\"u}hleisen} {and}
  \bibinfo{person}{Christian Bizer}.} \bibinfo{year}{2012}\natexlab{}.
\newblock \showarticletitle{Web {D}ata {C}ommons - extracting structured data
  from two large web corpora}. In \bibinfo{booktitle}{\emph{LDOW}}.
\newblock


\bibitem[\protect\citeauthoryear{Neumaier, Umbrich, and Polleres}{Neumaier
  et~al\mbox{.}}{2016}]%
        {neumaier2016automated}
\bibfield{author}{\bibinfo{person}{Sebastian Neumaier},
  \bibinfo{person}{J{\"u}rgen Umbrich}, {and} \bibinfo{person}{Axel Polleres}.}
  \bibinfo{year}{2016}\natexlab{}.
\newblock \showarticletitle{Automated quality assessment of metadata across
  open data portals}.
\newblock \bibinfo{journal}{\emph{Journal of Data and Information Quality
  (JDIQ)}} \bibinfo{volume}{8}, \bibinfo{number}{1} (\bibinfo{year}{2016}),
  \bibinfo{pages}{1--29}.
\newblock


\bibitem[\protect\citeauthoryear{Perkel}{Perkel}{2016}]%
        {perkel2016democratic}
\bibfield{author}{\bibinfo{person}{Jeffrey Perkel}.}
  \bibinfo{year}{2016}\natexlab{}.
\newblock \showarticletitle{Democratic databases: science on GitHub}.
\newblock \bibinfo{journal}{\emph{Nature News}} \bibinfo{volume}{538},
  \bibinfo{number}{7623} (\bibinfo{year}{2016}), \bibinfo{pages}{127}.
\newblock


\bibitem[\protect\citeauthoryear{Plotly}{Plotly}{2018}]%
        {plotlyfeed}
\bibfield{author}{\bibinfo{person}{Plotly}.} \bibinfo{year}{2018}\natexlab{}.
\newblock \bibinfo{title}{Plotly Community Feed}.
\newblock
\newblock
\urldef\tempurl%
\url{https://chart-studio.plotly.com/feed/}
\showURL{%
\tempurl}


\bibitem[\protect\citeauthoryear{Rabanser, G{\"u}nnemann, and Lipton}{Rabanser
  et~al\mbox{.}}{2019}]%
        {rabanser2019failing}
\bibfield{author}{\bibinfo{person}{Stephan Rabanser}, \bibinfo{person}{Stephan
  G{\"u}nnemann}, {and} \bibinfo{person}{Zachary~C Lipton}.}
  \bibinfo{year}{2019}\natexlab{}.
\newblock \showarticletitle{Failing Loudly: An Empirical Study of Methods for
  Detecting Dataset Shift}.
\newblock \bibinfo{journal}{\emph{33rd Conference on Neural Information
  Processing Systems}} (\bibinfo{year}{2019}).
\newblock


\bibitem[\protect\citeauthoryear{Raffel, Shazeer, Roberts, Lee, Narang, Matena,
  Zhou, Li, and Liu}{Raffel et~al\mbox{.}}{2020}]%
        {raffel2020t5}
\bibfield{author}{\bibinfo{person}{Colin Raffel}, \bibinfo{person}{Noam
  Shazeer}, \bibinfo{person}{Adam Roberts}, \bibinfo{person}{Katherine Lee},
  \bibinfo{person}{Sharan Narang}, \bibinfo{person}{Michael Matena},
  \bibinfo{person}{Yanqi Zhou}, \bibinfo{person}{Wei Li}, {and}
  \bibinfo{person}{Peter~J Liu}.} \bibinfo{year}{2020}\natexlab{}.
\newblock \showarticletitle{Exploring the Limits of Transfer Learning with a
  Unified Text-to-Text Transformer}.
\newblock \bibinfo{journal}{\emph{JMLR}} (\bibinfo{year}{2020}).
\newblock


\bibitem[\protect\citeauthoryear{Ritze and Bizer}{Ritze and Bizer}{2017}]%
        {ritze2017matching}
\bibfield{author}{\bibinfo{person}{Dominique Ritze} {and}
  \bibinfo{person}{Christian Bizer}.} \bibinfo{year}{2017}\natexlab{}.
\newblock \showarticletitle{Matching web tables to {DB}pedia -- a feature
  utility study}.
\newblock \bibinfo{journal}{\emph{{EDBT}}} \bibinfo{volume}{42},
  \bibinfo{number}{41} (\bibinfo{year}{2017}), \bibinfo{pages}{19}.
\newblock


\bibitem[\protect\citeauthoryear{Ritze, Lehmberg, and Bizer}{Ritze
  et~al\mbox{.}}{2021}]%
        {t2dv2}
\bibfield{author}{\bibinfo{person}{Dominique Ritze}, \bibinfo{person}{Oliver
  Lehmberg}, {and} \bibinfo{person}{Christian Bizer}.}
  \bibinfo{year}{2021}\natexlab{}.
\newblock \bibinfo{title}{{T2Dv2 Gold Standard for Matching Web Tables to
  DBpedia}}.
\newblock
\newblock
\urldef\tempurl%
\url{http://webdatacommons.org/webtables/goldstandardV2.html}
\showURL{%
\tempurl}
\newblock
\shownote{Accessed: 01-05-2021.}


\bibitem[\protect\citeauthoryear{Tenneson}{Tenneson}{2016}]%
        {tenneson2016csv}
\bibfield{author}{\bibinfo{person}{Jeni Tenneson}.}
  \bibinfo{year}{2016}\natexlab{}.
\newblock \bibinfo{title}{{CSV} on the web: {A} primer.}
\newblock
\newblock
\urldef\tempurl%
\url{http://www.w3.org/TR/2016/NOTE-tabular-data-primer-20160225/}
\showURL{%
\tempurl}


\bibitem[\protect\citeauthoryear{{U}niversity}{{U}niversity}{2010}]%
        {wordnet}
\bibfield{author}{\bibinfo{person}{Princeton {U}niversity}.}
  \bibinfo{year}{2010}\natexlab{}.
\newblock \bibinfo{title}{About {W}ord{N}et.}
\newblock
\newblock
\urldef\tempurl%
\url{https://wordnet.princeton.edu}
\showURL{%
\tempurl}


\bibitem[\protect\citeauthoryear{van~den Burg, Naz{\'a}bal, and Sutton}{van~den
  Burg et~al\mbox{.}}{2019}]%
        {van2019wrangling}
\bibfield{author}{\bibinfo{person}{Gerrit~JJ van~den Burg},
  \bibinfo{person}{Alfredo Naz{\'a}bal}, {and} \bibinfo{person}{Charles
  Sutton}.} \bibinfo{year}{2019}\natexlab{}.
\newblock \showarticletitle{Wrangling messy {CSV} files by detecting row and
  type patterns}.
\newblock \bibinfo{journal}{\emph{Data Mining and Knowledge Discovery}}
  \bibinfo{volume}{33}, \bibinfo{number}{6} (\bibinfo{year}{2019}),
  \bibinfo{pages}{1799--1820}.
\newblock


\bibitem[\protect\citeauthoryear{Viegas, Wattenberg, Van~Ham, Kriss, and
  McKeon}{Viegas et~al\mbox{.}}{2007}]%
        {viegas2007manyeyes}
\bibfield{author}{\bibinfo{person}{Fernanda~B Viegas}, \bibinfo{person}{Martin
  Wattenberg}, \bibinfo{person}{Frank Van~Ham}, \bibinfo{person}{Jesse Kriss},
  {and} \bibinfo{person}{Matt McKeon}.} \bibinfo{year}{2007}\natexlab{}.
\newblock \showarticletitle{Manyeyes: a site for visualization at internet
  scale}.
\newblock \bibinfo{journal}{\emph{IEEE transactions on visualization and
  computer graphics}} \bibinfo{volume}{13}, \bibinfo{number}{6}
  (\bibinfo{year}{2007}), \bibinfo{pages}{1121--1128}.
\newblock


\bibitem[\protect\citeauthoryear{Wang, Narayanan, and Russakovsky}{Wang
  et~al\mbox{.}}{2020}]%
        {wang2020revise}
\bibfield{author}{\bibinfo{person}{Angelina Wang}, \bibinfo{person}{Arvind
  Narayanan}, {and} \bibinfo{person}{Olga Russakovsky}.}
  \bibinfo{year}{2020}\natexlab{}.
\newblock \showarticletitle{REVISE: A tool for measuring and mitigating bias in
  visual datasets}. In \bibinfo{booktitle}{\emph{European Conference on
  Computer Vision}}. Springer, \bibinfo{pages}{733--751}.
\newblock


\bibitem[\protect\citeauthoryear{Wang, Shiralkar, Lockard, Huang, Dong, and
  Jiang}{Wang et~al\mbox{.}}{2021}]%
        {wang2021tcn}
\bibfield{author}{\bibinfo{person}{Daheng Wang}, \bibinfo{person}{Prashant
  Shiralkar}, \bibinfo{person}{Colin Lockard}, \bibinfo{person}{Binxuan Huang},
  \bibinfo{person}{Xin~Luna Dong}, {and} \bibinfo{person}{Meng Jiang}.}
  \bibinfo{year}{2021}\natexlab{}.
\newblock \showarticletitle{{TCN}: {T}able Convolutional Network for Web Table
  Interpretation}.
\newblock \bibinfo{journal}{\emph{arXiv preprint arXiv:2102.09460}}
  (\bibinfo{year}{2021}).
\newblock


\bibitem[\protect\citeauthoryear{Wang, Zhang, Chen, Jagadish, Ooi, and
  Tan}{Wang et~al\mbox{.}}{2016}]%
        {wang2016database}
\bibfield{author}{\bibinfo{person}{Wei Wang}, \bibinfo{person}{Meihui Zhang},
  \bibinfo{person}{Gang Chen}, \bibinfo{person}{HV Jagadish},
  \bibinfo{person}{Beng~Chin Ooi}, {and} \bibinfo{person}{Kian-Lee Tan}.}
  \bibinfo{year}{2016}\natexlab{}.
\newblock \showarticletitle{Database meets deep learning: Challenges and
  opportunities}.
\newblock \bibinfo{journal}{\emph{ACM SIGMOD Record}} \bibinfo{volume}{45},
  \bibinfo{number}{2} (\bibinfo{year}{2016}), \bibinfo{pages}{17--22}.
\newblock


\bibitem[\protect\citeauthoryear{WebDataCommons}{WebDataCommons}{2021}]%
        {webtables2012}
\bibfield{author}{\bibinfo{person}{WebDataCommons}.}
  \bibinfo{year}{2021}\natexlab{}.
\newblock \bibinfo{title}{WDC Web Table Corpus 2012}.
\newblock
\newblock
\urldef\tempurl%
\url{http://webdatacommons.org/webtables/2012/relationalStatistics.html}
\showURL{%
\tempurl}


\bibitem[\protect\citeauthoryear{Weikum}{Weikum}{2021}]%
        {weikum2021knowledge}
\bibfield{author}{\bibinfo{person}{Gerhard Weikum}.}
  \bibinfo{year}{2021}\natexlab{}.
\newblock \showarticletitle{Knowledge graphs 2021: a data odyssey}.
\newblock \bibinfo{journal}{\emph{Proceedings of the VLDB Endowment}}
  \bibinfo{volume}{14}, \bibinfo{number}{12} (\bibinfo{year}{2021}),
  \bibinfo{pages}{3233--3238}.
\newblock


\bibitem[\protect\citeauthoryear{Yin, Neubig, Yih, and Riedel}{Yin
  et~al\mbox{.}}{2020}]%
        {yin2020tabert}
\bibfield{author}{\bibinfo{person}{Pengcheng Yin}, \bibinfo{person}{Graham
  Neubig}, \bibinfo{person}{Wen-tau Yih}, {and} \bibinfo{person}{Sebastian
  Riedel}.} \bibinfo{year}{2020}\natexlab{}.
\newblock \showarticletitle{Ta{BERT}: Pretraining for Joint Understanding of
  Textual and Tabular Data}. In \bibinfo{booktitle}{\emph{ACL}}.
\newblock


\bibitem[\protect\citeauthoryear{Zhang and Balog}{Zhang and Balog}{2020}]%
        {zhang2020web}
\bibfield{author}{\bibinfo{person}{Shuo Zhang} {and} \bibinfo{person}{Krisztian
  Balog}.} \bibinfo{year}{2020}\natexlab{}.
\newblock \showarticletitle{Web table extraction, retrieval, and augmentation:
  A survey}.
\newblock \bibinfo{journal}{\emph{ACM Transactions on Intelligent Systems and
  Technology (TIST)}} \bibinfo{volume}{11}, \bibinfo{number}{2}
  (\bibinfo{year}{2020}), \bibinfo{pages}{1--35}.
\newblock


\end{thebibliography}
